\documentclass{aa}  

\usepackage{graphicx}
\usepackage{array}
\usepackage{comment}
\usepackage{subcaption}
\usepackage{lscape}
\usepackage{txfonts}
\usepackage[colorlinks=true, linkcolor=blue, citecolor=blue, filecolor=blue, urlcolor=blue]{hyperref}

\begin{document}

   \title{Candidate gravitationally lensed submillimeter galaxies in Herschel-ATLAS associated with WISE elliptical counterparts}
    
   \author{J.~A.~Cano\inst{1,2}
          \and
          J.~González-Nuevo\inst{1,2}
          \and
          L.~Bonavera\inst{1,2}
          \and
          M.~M.~Cueli\inst{3,4}
          \and
          T.~J.~L.~C.~Bakx\inst{5}
          \and
          J.~M.~Casas\inst{1,2,6,7}
          \and
          R.~Fernández-Fernández\inst{1,2}
          \and
          D.~Crespo\inst{1,2}
          }

   \institute{Departamento de Fisica, Universidad de Oviedo, C. Federico Garcia Lorca 18, 33007 Oviedo, Spain\\\email{canojuan@uniovi.es}\label{inst1}\and
            Instituto Universitario de Ciencias y Tecnologías Espaciales de Asturias (ICTEA), C. Independencia 13, 33004 Oviedo, Spain\label{inst2}\and
            SISSA, Via Bonomea 265, 34136 Trieste, Italy\label{inst3}\and
            IFPU - Institute for fundamental physics of the Universe, Via Beirut 2, 34014 Trieste, Italy\label{inst4}\and
            Department of Space, Earth, \& Environment, Chalmers University of Technology, Chalmersplatsen 4, Gothenburg SE-412 96, Sweden\label{inst5}\and
            Instituto de Astrofísica de Canarias, E-38200 La Laguna, Tenerife, Spain\label{inst6}\and
            Universidad de La Laguna, Departamento de Astrofísica, E-38206 La Laguna, Tenerife, Spain\label{inst7}
             }
             
   \date{Received 15 May 2025; accepted 5 August 2025} 

  \abstract
   {}
   {We present a new and independent methodology for identifying gravitational lens candidates using data from the H-ATLAS and AllWISE surveys. Unlike previous approaches, which are typically biased toward bright, strongly lensed submillimeter galaxies (SMGs), our method uncovers fainter systems with lower magnifications. This enables the identification and individual study of lensing events that would otherwise only be accessible through statistical weak lensing analyses.
}
   {Our approach focuses on high-redshift SMGs from H-ATLAS in the range $1.2 < z < 4.0$, and searches for associated AllWISE sources within an angular distance of 18 arcsec. Candidate lenses are selected based on their WISE colors ($0.5 < \mathrm{W2} - \mathrm{W3} < 1.5$ mag), consistent with those of elliptical galaxies, and further filtered using $J-\mathrm{W1}$ color and photometric redshift cuts to reduce stellar contamination. This conservative selection yields 68 new lens candidates. We then performed SED fitting with CIGALE across UV to submillimeter wavelengths to estimate the physical properties of both the lenses and the background SMGs, and to assess the lensing nature of these candidates.
}
   {Despite the uncertainties, we were able to constrain key parameters such as stellar and dust masses, infrared luminosities, and star formation rates. In addition, the estimated magnifications for most candidates are modest, consistent with the weak lensing regime ($\mu \simeq 1{-}1.5$), although a few sources may require more precise modeling. Future efforts could refine this methodology to recover potential candidates outside our selection, and high-resolution follow-up observations will be essential to confirm the lensing nature of these sources and to further investigate their physical properties.
}
   {}

   \keywords{lensing --
                Galaxies: high-redshift --
                Submillimeter: galaxies --
                Gravitational lensing: strong --
                Gravitational lensing: weak
               }

  \titlerunning{Lensed SMG candidates in H-ATLAS with WISE counterparts}
  \authorrunning{J. A. Cano et al.}

\maketitle

\section{Introduction}\label{sec:intro}
In the past decades, submillimeter observations have revolutionized our understanding on the formation and evolution of galaxies by revealing an unexpected population of high-redshift dust-obscured galaxies known as submillimeter galaxies (SMGs). These galaxies, which have an extremely high star formation rate (SFR $\gtrsim 1000\ M_\odot\, \text{yr}^{-1}$; \citealp{Blain1999}), are in an unsustainable stage of evolution driven by violent feedback processes \citep{Andrews2011,Rowan-Robinson2016}, and provide the perfect scenario for testing existing models of galaxy formation (see \citealp{Casey2014} for a review). Moreover, thanks to their redshift distribution \citep[$z\gtrsim 1$;][]{Chapman2004,Chapman2005,Amblard2010,Lapi2011,GON12,Pearson2013}, the steepness of the number of counts ($\beta\gtrsim 3$), and the attenuation of stellar emission by dust, they are also ideal sources for lensing magnification studies, as shown by \cite{Negrello2007}. Gravitational lensing offers a unique opportunity to study these dusty star-forming galaxies at high resolution, but large samples of such lensed SMGs are needed to constrain evolutionary models and study their general properties in a statistical sense.

The sensitivity and spectral coverage of the Herschel Space Observatory \citep{Pilbratt2010} in the submillimeter, along with that of the South Pole Telescope (SPT; \citealt{Carlstrom2011}) and Planck at millimeter wavelengths, enabled the discovery of a substantial population of strongly lensed sources. Upon closer examination, many of these sources were confirmed as gravitationally lensed through ground-based and space-based observations (see \citealp{Negrello2010, Bussmann2012, Fu2012, Bussmann2013, Calanog2014, Messias2014, Negrello2014, Negrello2017} for Herschel; \citealp{Vieira2013, Hezaveh2013, Spilker2016}, for SPT; or \citealp{Canameras2015, Harrington2016} for Planck). During the Science Demonstration Phase of the Herschel Astrophysical Terahertz Large Area Survey \citep[H-ATLAS;][]{EAL10,Smith2017}, five of these earliest identified sources were reported by \citet{Negrello2010} using a simple flux density cut at 500\,$\mu$m ($S_{500} > 100$\,mJy). This approach ultimately led to a sample of 80 confirmed gravitational lenses using the full Herschel catalog \citep{Negrello2007,Negrello2017}. Following a similar approach, \cite{Wardlow2013} and \cite{Nayyeri2016} identified 11 and 77 additional strongly lensed galaxies, respectively, within the Herschel Multi-tiered Extragalactic Survey \citep[HerMES;][]{Oliver2012}, the HerMES Large Mode Survey (HeLMS), and the Herschel Stripe 82 Survey \citep[HerS;][]{Viero2014} fields. More recently, \cite{Bakx2020} extended this method to sources with VIKING \citep{VIKING2013} counterparts and $S_{500}>80$ mJy in the Herschel Bright Sources \citep[HerBS;][]{Bakx2018} catalog, while \cite{Bakx2024} refined it to include even fainter sources in the $15{-}85$ mJy range, successfully identifying 47 new strongly lensed SMGs in the GAMA-12 field of H-ATLAS. If extended to the rest of the H-ATLAS fields, this refined methodology would enable the identification of $\sim 3000$ lensed sources \citep{Bakx2024}.

In addition to flux density-based selection methods, other approaches have also been tested for identifying gravitational lenses. These include, for example, the use of the CO linewidth to luminosity relation in submillimeter spectroscopic observations \citep{Harris2012, Bothwell2013, Aravena2016, Neri2020} and statistical correlations based on various observables, such as redshifts, spatial distributions, luminosity percentiles, and the ratio of optical to submillimeter flux densities \citep{GON12,GON19}. While CO linewidth data have been shown to be unreliable for distinguishing lensed galaxies, especially at low magnifications \citep{Aravena2016}, the latter approach was used to compile a catalog of more than 1000 proposed lens candidates in H-ATLAS.

However, the most reliable method to confirm individual strong gravitational lenses is through direct observation of typical strong lensing features, such as arcs or Einstein rings, using high-resolution imaging. This can be done with optical and infrared space telescopes like HST, JWST, and Euclid \citep[e.g.,][]{Calanog2014, EuclidCollaboration2025, Nightingale2025}; millimeter and submillimeter interferometers such as ALMA and SMA \citep[e.g.,][]{Bussmann2012, Bussmann2013, Dong2019, Bakx2024}; or ground-based optical and near-infrared (NIR) telescopes equipped with adaptive optics, like the Keck Observatory \citep[e.g.,][]{Calanog2014}. 

Although this technique is highly effective, conducting such observations across wide-field surveys is impractical due to the significant time and resources required. As a result, high-resolution follow-up studies are typically performed on smaller, previously selected samples of gravitational lens candidates identified through other methods. These targeted observations not only confirm the lensing nature of the sources but also provide detailed information about their physical properties and the mass distribution of the foreground lenses \citep[e.g.,][]{Negrello2010, Bussmann2013, Calanog2014, Dye2018, Borsato2024}.

In this paper we present a new method for selecting gravitational lens candidates within the H-ATLAS and AllWISE surveys. We focused on high-redshift SMGs from H-ATLAS whose mid-infrared (MIR) emission deviates from that expected for starburst galaxies and may instead originate from foreground ellipticals, potentially acting as gravitational lenses. These candidates were identified through a color-color selection in WISE, followed by additional filtering to reduce stellar contamination. Unlike previous catalogs, this method is not biased toward high submillimeter flux densities, making it better suited to explore weaker gravitational lenses at lower magnifications. Furthermore, we performed spectral energy distribution (SED) analyses from the ultraviolet (UV) to the submillimeter, aiming to derive the main physical parameters of both the lensing and lensed galaxies, and to estimate the magnification of the background sources. This approach allows us to assess the likelihood of a lensing scenario, although high-resolution imaging is still required to confirm the lensing nature of these candidates.

This article is organized as follows. Section \ref{sec:methodology} outlines the candidate selection methodology and the datasets used. The SED fitting procedure is presented in Sect. \ref{sec:SED_fitting}. In Sect. \ref{sec:validation} we validate the method by addressing stellar contamination, identifying the most reliable candidates, and comparing our results with existing catalogs. The results of the SED analysis are discussed in Sect. \ref{sec:discussion}, and the main conclusions are summarized in Sect. \ref{sec:conclusions}. In addition, Appendix \ref{sec:mock_analysis} includes the results from mock analyses, while Appendix \ref{sec:Theoretical_framework} offers a brief theoretical description of the lensing magnification produced by a singular isothermal sphere density profile. Finally, Appendix \ref{sec:SED_models} includes the full set of best-fit models, and Appendix \ref{sec:full_catalog} provides the complete catalog along with some of their most relevant physical properties. Throughout this paper, we assume a flat $\Lambda$ cold dark matter cosmology with the best-fitting parameters derived from the Planck results \citep{Planck2020}, which are $\Omega_m = 0.315$, $\Omega_\Lambda = 0.685$, and $h = 0.674$.

\section{Selection methodology}\label{sec:methodology}
\subsection{H-ATLAS}\label{sec:H-ATLAS}
H-ATLAS \citep{EAL10,Smith2017} is the largest area survey carried out by the Herschel Space Observatory \citep{Pilbratt2010}, covering $\sim 610$ deg$^2$ with its two instruments, PACS \citep{Poglitsch2010} and SPIRE \citep{Griffin2010}, operating in five photometric bands between 100 and 500 $\mu$m. The survey comprises five different fields, three of which are located on the celestial equator (known as the GAMA fields, or G09, G12, and G15; \citealp{Ibar2010, Pascale2011, Rigby2011, Bourne2016, Valiante2016}, as they coincide with those from the GAMA mission; \citealp{DRI11}), and the other two are located at the North and South Galactic poles \citep[NGP and SGP;][]{Smith2017, Furlanetto2018, Maddox2018, Ward2022}. Photometric redshifts of these sources were independently estimated by \cite{GON19} (and references therein) based on a minimum $\chi^2$ fit to SPIRE and PACS data, using the SED template of SMM J2135-0102 (commonly known as the Cosmic Eyelash) at $z=2.3$ \citep{Ivison2010,Swinbank2010}. Compared to other templates and methods, this galaxy has proven to give very accurate results for photometric redshift calculations at submillimeter wavelengths, at least for $z\gtrsim 0.8$, with a mean precision of $\Delta z/(1 + z) \simeq 0.1$ when compared with available spectroscopic measurements \citep{Ivison2016}.

Subsequently, we used these photometric redshift estimations to select a sample of SMGs in the H-ATLAS catalog, limiting them between $1.2<z<4.0$ for higher reliability. In addition, we also required our sources to have at least $4\sigma$ and $3\sigma$ detections at 250 and 350 $\mu$m, respectively, in order to get a cleaner sample. This procedure ends with a sample of $102\,212$ SMGs, with a median redshift of $\sim 2.2$.

\subsection{AllWISE}\label{sec:WISE}
AllWISE is the compiled All-Sky Data Release of the Wide-field Infrared Survey Explorer \citep[WISE;][]{Wright2010} mission, which observed the entire sky in four photometric bands centered at 3.4, 4.6, 12, and 22 $\mu$m, and denoted as W1, W2, W3, and W4, respectively. Using the cross-matching tool in TOPCAT,\footnote{\url{https://www.star.bris.ac.uk/\%7Embt/topcat/}} we searched for MIR counterparts of these SMGs with the aim of identifying candidates whose emission deviates from that expected for a starburst galaxy, and may instead originate from an elliptical galaxy acting as a gravitational lens. This was done by matching AllWISE sources to the Herschel positions within a radius of 18 arcsec, which is approximately the spatial resolution of the SPIRE 250 $\mu$m band (the most precise of the five Herschel channels). The angular resolution of WISE is significantly better, about $\sim 6$ arcsec at 3.4 $\mu$m, so we considered the former to minimize the loss of possible candidates.

Other studies considered instead a radius of only 10 arcsec to search for Herschel counterparts \citep{Smith2011, Bond2012, Bourne2016, Furlanetto2018}, as the majority of true AllWISE matches ($>99\%$) fall within that distance from Herschel positions. However, this is not true for the study of weak lensing, which can occur even at much larger angular separations. In contrast, strong lensing occurs at smaller scales, typically about $1{-}2$ arcsec, which translates to a distance of about $5{-}10$ arcsec when taking into account astrometric uncertainties. Therefore, a value of 18 arcsec seems to be a good compromise between the loss of possible candidates and contamination with spurious sources, while extending the search for lens candidates to larger angular scales and thus fainter magnifications.

The described procedure leads to the identification of $97\,608$ best matches in the AllWISE catalog. In the presence of multiple counterparts, we considered the closest to the Herschel coordinates as the most probable match, although more sophisticated techniques can be employed. One of the most common is the likelihood ratio \citep{Richter1975,Sutherland1992,Ciliegi2003} method, used to identify submillimeter counterparts in the Sloan Digital Sky Survey \citep{Smith2011} or WISE \citep{Bond2012}, but this kind of analysis is beyond the scope of this study. However, adopting such a methodology would not significantly affect our results, as it tends to be less effective for sources with larger angular separations or low magnifications. More importantly, it would not mitigate stellar contamination, which, as discussed in Sect. \ref{sec:star_removal}, represents the primary source of cross-match uncertainty in our analysis.

\begin{figure}
\centering
\includegraphics[width=\linewidth]{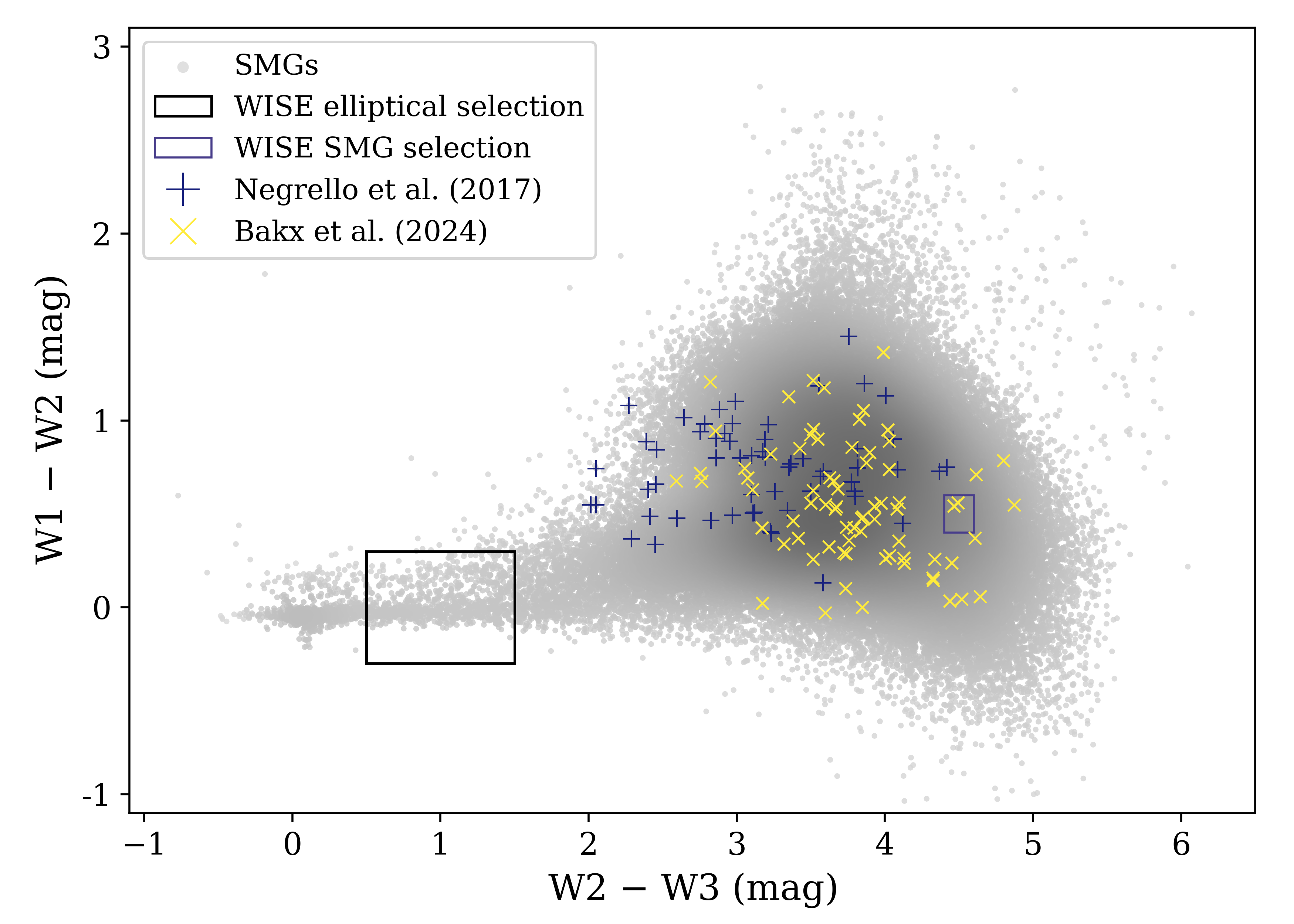}
\caption{Color-color diagram based on WISE magnitudes (in the Vega system), showing the positions of SMGs from H-ATLAS (in gray), and the elliptical galaxy selection (black box), defined by $0.5 < \mathrm{W2}-\mathrm{W3} < 1.5$ and $|\mathrm{W1}-\mathrm{W2}|<0.3$ mag. The blue box corresponds to a sample of WISE-selected starburst galaxies used to test our fitting and cross-matching procedure. Confirmed lens candidates from the catalogs of \citet{Negrello2017} and \citet{Bakx2024} are shown as plus signs and crosses, respectively.}
\label{fig:EllipticalsW12 vs W23}
\end{figure}

The key aspect about WISE magnitudes and colors is that they enable us to classify these sources into different types of astronomical objects using a color-color diagram, as shown first by \cite{Wright2010} and subsequently discussed in \cite{Nikutta2014}. We can indeed compare our Fig. \ref{fig:EllipticalsW12 vs W23} with Fig. 12 from \cite{Wright2010} to classify our WISE counterparts, proving that most of the sources fall within the region associated with different types of starburst galaxies, such as SMGs, ultra-luminous infrared galaxies (ULIRGs), luminous infrared galaxies (LIRGs), or low-ionization nuclear emission-line regions (LINERs). All of them are consistent with the initial redshift selection in H-ATLAS, but there is a significant portion of sources lying in the zone related to elliptical galaxies, which are typically found at low redshifts, exhibit little star formation, and have either consumed or expelled most of the dust they contained, making them entirely distinct from the SMG population. In view of these results, there are three possible explanations for this apparent contradiction: (i) those objects are actually the same, and it would mean that there is some issue in the Herschel or WISE data for them; (ii) those objects are different, but they have been identified as the same due to the randomness of their position and proximity (i.e., due to cross-match errors and the limited angular resolution of the telescopes); (iii) those objects are different and form a gravitational lens, magnifying the brightness of the SMG.

These objects thus appear as potential candidates to be strong or weak gravitational lenses, provided their identification is accurate and the elliptical galaxy is sufficiently massive to act as the lens. Indeed, as shown in the top panel of Fig. \ref{fig:angDist}, the angular separation distribution of the MIR counterparts relative to the submillimeter sources strongly supports the scenario in which two physically unrelated objects are aligned along the line of sight, although this does not necessarily imply the presence of a gravitational lens. 

By convention, for this work, elliptical galaxies were selected within the $0.5 < \mathrm{W2} - \mathrm{W3} < 1.5$ magnitude range (Fig. \ref{fig:EllipticalsW12 vs W23}). We also imposed the condition $|\mathrm{W1}-\mathrm{W2}|<0.3$ mag to exclude sources with highly scattered colors, as this threshold corresponds to approximately $1\sigma$ in that color.

We note, however, that stars share similar WISE colors with ellipticals, so they must be faithfully removed from the sample in some other way. While it is unlikely that stars are detected by Herschel, unless they are either dusty pre-main-sequence or post-asymptotic giant branch stars, or possess debris disks, they can be easily seen at shorter wavelengths and contaminate the lens sample. This kind of separation between stars and galaxies is more straightforward in optical and NIR wavelengths, where their colors are notably different, but several approaches can be employed. A detailed description of the procedures used and further discussion on this topic is provided in Sect. \ref{sec:star_removal}. However, following \cite{Nikutta2014}, some other color and magnitude cuts can be employed to further reduce stellar contamination at this point, with the use of WISE data only. For instance, we require our candidates to have $\mathrm{W1}>12$ mag, $\mathrm{W3}>10$ mag, and $\mathrm{W3}-\mathrm{W4}>2$ mag. This way we get an initial sample of 849 WISE elliptical candidates (representing 0.87\% of the total number of counterparts), although a high level of stellar contamination is still present.

\begin{table}
\caption[]{Data used for SED fitting and analysis, comprising a total of 21 photometric bands from the UV to submillimeter.}\label{tab:filters}
    \small
     $
         \begin{array}{p{0.3\linewidth}p{0.25\linewidth}l}
            \hline
            \noalign{\smallskip}
            Survey & Coverage & \text{Filters} \\
            \noalign{\smallskip}
            \hline
            \noalign{\smallskip}
            GALEX & All patches & FUV, NUV \\
            SDSS & G09-15, NGP & u, g, r, i, z\\
            VST-ATLAS & SGP & u, g, r, i, z\\
            Pan-STARRS1 & All patches & g, r, i, z, y\\
            VISTA-VIKING & G09-15, SGP & J, H, K_s \\
            UKIRT & G09-15, NGP & J, H, K \\
            2MASS & All patches & J, H, K_s \\
            WISE & All patches & 3.4, 4.6, 12, 22\ (\mu\mathrm{m})\\
            Herschel/PACS & All patches & 100, 160\ (\mu\mathrm{m})\\
            Herschel/SPIRE & All patches & 250, 350, 500\ (\mu\mathrm{m})\\
            \noalign{\smallskip}
            \hline
         \end{array}
     $
\end{table}

\subsection{Ancillary data}
In order to build up the SED of these candidates and study their physical properties from a panchromatic perspective, we utilized data from several catalogs covering additional wavelengths from the UV to NIR, as summarized in Table \ref{tab:filters}. These catalogs, from shorter to longer wavelengths, encompass the Galaxy Evolution Explorer GR6/7 \citep[GALEX;][]{GALEX2014}, the Sloan Digital Sky Survey DR16 \citep[SDSS;][]{SDSSDR16}, the VLT Survey Telescope ATLAS \citep{ATLAS2015}, the Panoramic Survey Telescope and Rapid Response System DR1 \citep[Pan-STARRS;][]{panstarrs}, the VISTA Kilo-degree Infrared Galaxy Survey DR2 \citep[VIKING;][]{VIKING2013}, the UKIRT Infrared Deep Sky Survey DR9 \citep[UKIDSS;][]{UKIDSS2007}, and the Two Micron All Sky Survey \citep[2MASS;][]{2MASS2006}. Photometric data from these surveys were first converted to the AB system, and then extinction-corrected using the maps from \citet{Schlegel1998} and extinction coefficients from \citet{Yuan2013}.

However, when multiple datasets are available for the same object and filter, we selected, for each source, the datasets that provided the most consistent flux densities between consecutive wavelength bands. This approach minimizes discontinuities in the SED, specifically between the W1 and J bands, and between J and z, hence reducing source mismatches with nearby counterparts. Despite these efforts, small inconsistencies persist in the optical bands for some candidates, as shown in Appendix \ref{sec:SED_models}.

On the other hand, cross-matching data from different surveys presents additional challenges, as differences in angular resolution and wavelength coverage can complicate the unambiguous identification of counterparts. This is the case of Herschel and WISE when compared to optical and NIR surveys, as submillimeter and MIR sources with lower resolution can be resolved into multiple components at shorter wavelengths. However, by combining WISE positions with Herschel data, the positional accuracy of the SMGs can be significantly improved. The same applies to elliptical galaxies, which are often invisible in the submillimeter but can be detected by WISE. Consequently, WISE coordinates generally provide more reliable matches than H-ATLAS data alone, owing to the better angular resolution and wavelength proximity to optical surveys.

\begin{figure}
\centering
\includegraphics[width=0.82\linewidth]{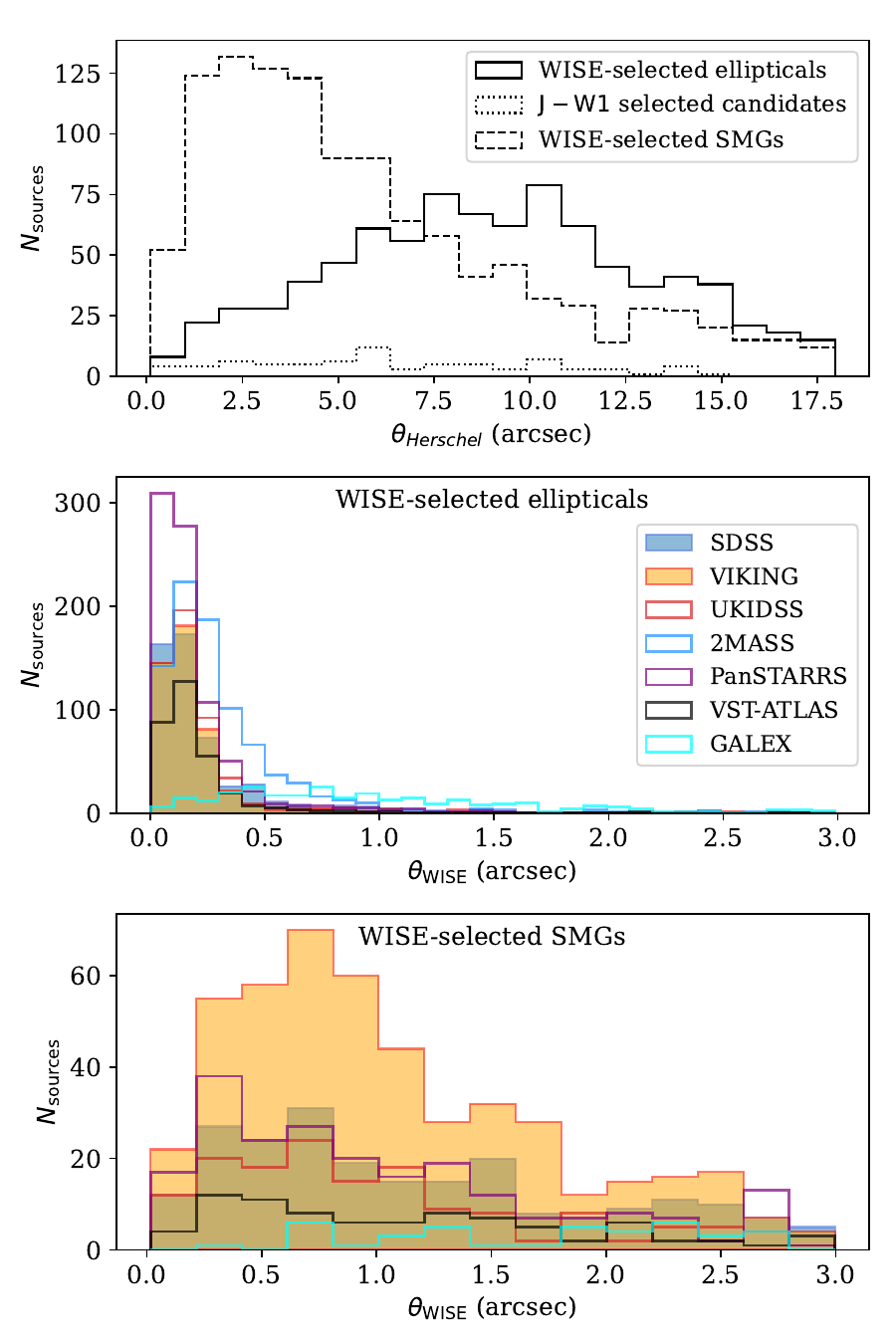}
\caption{Angular separations of the AllWISE lenses and their best counterparts in other surveys with respect to their parent catalog positions (i.e., Herschel and WISE, respectively for the top and central panels). For comparison, the top and bottom panels show the angular separations of the subset of WISE-selected SMGs, which are more likely to be true counterparts of the Herschel sources.}
\label{fig:angDist}
\end{figure}

In our case, we cross-matched the WISE lens candidates by selecting the best match inside a search radius of 3 arcsec. As shown in the middle panel of Fig. \ref{fig:angDist}, most optical and NIR counterparts lie within 1 arcsec of the WISE positions, indicating reliable associations consistent with the astrometric uncertainties of the data.

\subsection{Independent sample of WISE-selected SMGs}
To compare and test our results, we also selected an independent random sample of 1100 unlensed SMGs (i.e., not selected based on lensing criteria) using the WISE colors plotted in Fig. \ref{fig:EllipticalsW12 vs W23}, located within a $0.2$ mag square centered on $\mathrm{W2}-\mathrm{W3}=4.5$ and $\mathrm{W1}-\mathrm{W2}=0.5$. This region is typically associated with ULIRGs and SMGs, so contamination from other objects is expected to be lower than in our lens sample, making their optical and NIR counterparts more likely to correspond to the submillimeter emission. Their WISE counterparts are generally closer than those of the lenses (top panel of Fig. \ref{fig:angDist}), while the optical counterparts are usually more distant, though still within astrometric uncertainties (bottom panel). Therefore, they can serve as an independent sample to validate our analysis method and look for any possible issues that may affect our results, such as cross-matching errors or SED fitting degeneracies. However, due to their higher dust attenuation and higher redshifts compared to the lenses, optical or NIR detections are only available for fewer than 300 galaxies.

Although this implies a strong bias in the population of these galaxies, those lacking optical and NIR data were excluded from the analysis to minimize uncertainties in the estimated properties (see Sect. \ref{sec:non-lensed_validation}). We must therefore be cautious when comparing those galaxies with our lensed candidates, as the majority of SMGs are invisible at such wavelengths due to their strong dust attenuation. This could lead to significant differences in their physical properties, as discussed in Sect. \ref{sec:comparison_SMGs}.

\begin{table*}[t]
\centering
\caption{Grids of input parameters used in CIGALE for the best-fit model computation of both the lens and submillimeter components of a SED.}\label{tab:CIGALE_Param_region1}
\small
\begin{tabular}{p{0.45\linewidth}>{\centering\arraybackslash}p{0.24\linewidth}>{\centering\arraybackslash}p{0.24\linewidth}}
\hline\hline
\noalign{\smallskip}
\textbf{Parameter} & \textbf{Lens component} & \textbf{SMG component}\\
\noalign{\smallskip}
\hline
\noalign{\smallskip}
\textbf{SFH} & delayedbq \citep{Ciesla2017} & delayedbq \citep{Ciesla2017}\\
\noalign{\smallskip}
\hline
\noalign{\smallskip}
e-folding time of the main stellar population, $\tau_{\text{main}}$ (Myr) & 200, 500, 750, 1000, 1500 & 2000, 3000, 5000\\
Age of the main stellar population, $t_{\text{main}}$ (Myr) & 5000, 6000, 7000, 8000, 9000, 10000 & 1000, 1500, 2000, 3000, 4000\\
Age of the late burst/quench episode, $t_{\text{bq}}$ (Myr) & 200 & 10, 20, 100\\
Ratio of the SFR after/before $t_{\text{bq}}$, $r_{\text{SFR}}$ & 0.01, 0.1, 0.3 & 1.5, 3, 10, 20, 30\\
\noalign{\smallskip}
\hline\hline
\noalign{\smallskip}
\textbf{Stellar population model} & \cite{BC2003} & \cite{BC2003}\\
\noalign{\smallskip}
\hline
\noalign{\smallskip}
IMF & \cite{Chabrier2003} & \cite{Chabrier2003}\\
Metallicity ($Z_\odot$) & 0.02 & 0.008\\
\noalign{\smallskip}
\hline\hline
\noalign{\smallskip}
\textbf{Dust attenuation} & \cite{Calzetti2000} & \cite{Calzetti2000}\\
\noalign{\smallskip}
\hline
\noalign{\smallskip}
Color excess of the nebular lines light, $E(B-V)_{\text{lines}}$ & 0.001, 0.003, 0.0075 & 1.0, 1.2, 1.4, 1.6\\
$E(B-V)_{\text{old}}/E(B-V)_{\text{young}}$ & 0.25, 0.50, 0.75 & 0.25, 0.5, 0.75\\
Slope of the power law modifying the attenuation curve, $\delta$ & -1.0, -0.5, -0.25, 0.0 & 0.0\\
\noalign{\smallskip}
\hline\hline
\noalign{\smallskip}
\textbf{Dust grain model} & \cite{Draine2014} & \cite{Draine2014}\\
\noalign{\smallskip}
\hline
\noalign{\smallskip}
Mass fraction of PAHs, $q_{\text{PAH}}$ (\%) & 0.47, 1.12 & 1.12, 3.19, 5.26, 7.32\\
Minimum radiation field, $U_{\text{min}}$ & 20 & 10, 20, 25, 30\\
Power-law slope $\text{d}U/\text{d}M$, $U^\alpha$ & 2.0 & 2.0\\
Fraction illuminated from $U_\text{min}$ to $U_\text{max}$, $\gamma$ & 0.01 & 0.02\\
\noalign{\smallskip}
\hline\hline
\noalign{\smallskip}
\textbf{AGN model} & -- & \cite{Fritz2006}\\
\noalign{\smallskip}
\hline
\noalign{\smallskip}
Ratio of the maximum to minimum radii of the dust torus & -- & 60\\
Optical depth at 9.7$\mu$m, $\tau_{9.7}$ & -- & 3.0\\
$\beta$ & -- & -0.5\\
$\gamma$ & -- & 0.0\\
Opening angle of the dust torus (deg) & -- & 100\\
Angle between equatorial axis and line of sight (deg) & -- & 10.1, 60.1\\
$\delta$ & -- & -0.36\\
AGN fraction & -- & 0.0, 10, 30, 50\\
E(B-V) & -- & 0.03\\
Temperature of the polar dust (K) & -- & 100\\
Emissivity index of the polar dust & -- & 1.6\\
\noalign{\smallskip}
\hline\hline
\noalign{\smallskip}
\textbf{Redshift} & $z=0.01$ to 1.0, with $\Delta z=0.01$ & $z=1.0$ to 4.5, with $\Delta z=0.1$\\
\noalign{\smallskip}
\hline
\end{tabular}
\tablefoot{This configuration gives us a set of $648\,000$ models for the lenses and $1\,555\,200$ for the SMGs. Moreover, an additional set of $53\,222\,400$ parameters was created to perform SED analysis over the entire SED range by combining values from the individual components, and an optional AGN component was added for an additional run of the unlensed SMGs.}
\end{table*}

\section{SED fitting}\label{sec:SED_fitting}
For the purpose of this work, we made use of the SED fitting code CIGALE \citep[see][and references therein]{Boquien2019} to model the SEDs of both the lenses and SMGs in our sample and derive their main physical properties. CIGALE serves as a versatile SED-modeling tool, facilitating the construction of the SED of a galaxy across the UV to submillimeter spectrum, and assuming energy conservation between the energy absorbed by dust and that emitted by stars. It has a modular design, allowing the user to combine several models and libraries that account for each different emission processes contributing to the SED. Then, it employs a Bayesian approach to find the set of parameters that best represents the observed SED. In our case, the stellar radiation field was built based on the \cite{BC2003} population synthesis model and the \cite{Chabrier2003} initial mass function (IMF). Following \cite{Ciesla2016,Ciesla2017}, we adopted a delayed star formation history (SFH) for the main stellar population, paired with an additional constant burst or quenching episode during the last 200 Myr. Stellar emission from both populations was then attenuated by dust using a modified \cite{Calzetti2000} attenuation law for starburst galaxies, and subsequently re-emitted through thermal radiation at IR wavelengths with the \cite{Draine2014} dust grain model. For simplicity and consistency during analysis, the same models were used for both the lenses and submillimeter galaxies, although with different sets of parameters to adapt to the nature of each source type (see Table \ref{tab:CIGALE_Param_region1} for a complete description of the parameters of each model and their values). We also tested the addition of a small active galactic nucleus (AGN) component to the sample of unlensed SMGs, for comparison purposes, but it was omitted for the fit of the lensed candidates. This was accomplished with the \cite{Fritz2006} model with two obscured AGN scenarios, each one with different AGN fractions contributing between $0{-}50\%$ to the total observed infrared emission.

\begin{figure}
    \centering
    \includegraphics[width=0.9\linewidth]{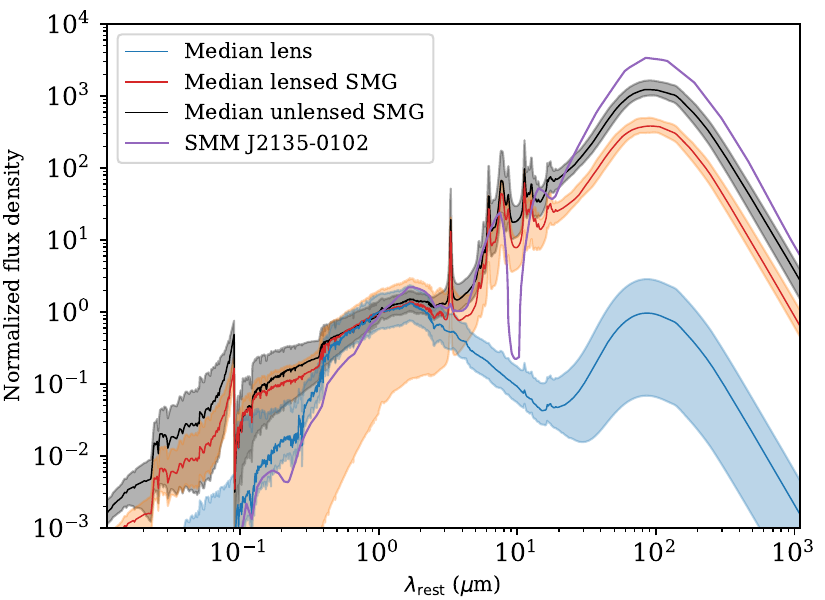}
    \caption{Rest-frame composition of the best-fit models created with CIGALE, normalized at $\lambda=1\ \mu$m. The solid blue, red, and black curves represent the median SED models of the lens candidates, their lensed SMGs, and the sample of unlensed SMGs, respectively, while the shaded areas indicate the range between the 16th and 84th percentiles to the median values. We note that the lensed sources are somewhat fainter than the unlensed ones. Additionally, the SED template of SMM J2135-0102 \citep{Ivison2010,Swinbank2010} is plotted in purple for comparison purposes.}
    \label{fig:Median_SEDs}
\end{figure}

With the appropriate models and input parameters, CIGALE can accurately model the SED of any galaxy, but in the case of a lensing, event two galaxies are involved: the foreground galaxy (lens) and the background galaxy (lensed SMG). Therefore, we needed to run CIGALE separately for the two components of the SEDs to take into account their two distinct contributions (Fig. \ref{fig:Median_SEDs}), and decide which filters are best used to model each component.

From the 21 filters included in our catalog (Table \ref{tab:filters}), the first 15 located at shorter wavelengths (i.e., FUV, NUV, u, g, r, i, z, y, J, H, K, Ks, W1, W2, and W3) were considered to account for the lens component of the SEDs, covering the near-UV to mid-IR range. Meanwhile, W4, along with the five PACS and SPIRE bands, were used to characterize the dust emission component of the SMGs. This implies that the SEDs of both galaxies were modeled exclusively using the emission information contained in these ranges. In addition, the W3 and W4 bands show visible traces of blending in some cases, so they may have a contribution from both galaxies. Since W4 predominantly accounts for the emission of the SMGs, it was assigned to them rather than the ellipticals to minimize its impact on the dust emission component of the latter, and better constrain the mid-IR emission and stellar parameters of the SMGs. On the other hand, W3 was assigned to the lens component, as the contamination from the SMGs is expected to be lower in that filter.

\subsection{Full SED model fits}
In addition to the two-component analysis, a global fit was performed using all available filters to discard the possibility of a single galaxy being consistent with all of them. In most cases, there is a significant mismatch between the optical and submillimeter fluxes with respect to the global best-fit model, and in fact no candidates show a single SED compatible with all the data. This is primarily a consequence of the high-redshift selection of submillimeter sources, whose dust emission peaks are highly inconsistent with those of lower-redshift galaxies. Consequently, two galaxies at different redshifts are needed to explain the observations. However, it is important to note that even if the model of a single galaxy agreed well with the data, this would not entirely rule out the existence of the lensing phenomenon, since the SED of two different galaxies could, in some cases, mimic that of a single galaxy, particularly when fitting broadband data. Conversely, the presence of two galaxies along the line of sight at different redshifts does not necessarily imply the existence of a strong gravitational lens.

\subsection{AGN component}
SMGs radiate most of their energy at submillimeter wavelengths due to their high star formation rates and dust content, but some fraction of the photons may originate in the accretion disk and torus of an obscured AGN, as supported by both observations \citep[e.g.,][]{Serjeant2010, Johnson2013, Symeonidis2013, Chen2020} and hydrodynamic simulations \citep[e.g.,][]{Hopkins2008,Hobbs2011,McAlpine2019}. However, the fraction of AGN that is present among the SMGs and the actual AGN contribution to the infrared luminosity are still unclear. While including an AGN component in the models may seem a reasonable strategy, the reduced spectral coverage for the lensed SMGs and the large number of degrees of freedom of the models would unnecessarily increase the complexity of the analysis without providing any significant constraints. Conversely, the situation is different for the sample of unlensed SMGs, where NIR and optical data are available. This allows us to test the inclusion of a small AGN component and investigate any potential biases it might introduce compared to models without it.

\section{Validation}\label{sec:validation}
\subsection{Star/galaxy separation}\label{sec:star_removal}
As mentioned in Sect. \ref{sec:WISE}, stars share very similar WISE colors with elliptical galaxies, and thus a significant number of stars are expected to be present in our sample due to the intrinsic nature of the selection procedure. Therefore, removing these interlopers is a crucial step in identifying true lens candidates. Although, in principle, the presence of a nearby star in the field should not affect the lensing effect produced by a galaxy, confusing lenses with local stars would contaminate the sample with chance alignments of stars and SMGs, reducing the efficiency of our method in selecting gravitational lenses. We therefore aim to exclude those candidates where the closest WISE counterpart to the H-ATLAS position resembles a star in optical wavelengths, ensuring that WISE and optical data are related to the same source.

\begin{figure}
    \centering
    \includegraphics[width=\linewidth]{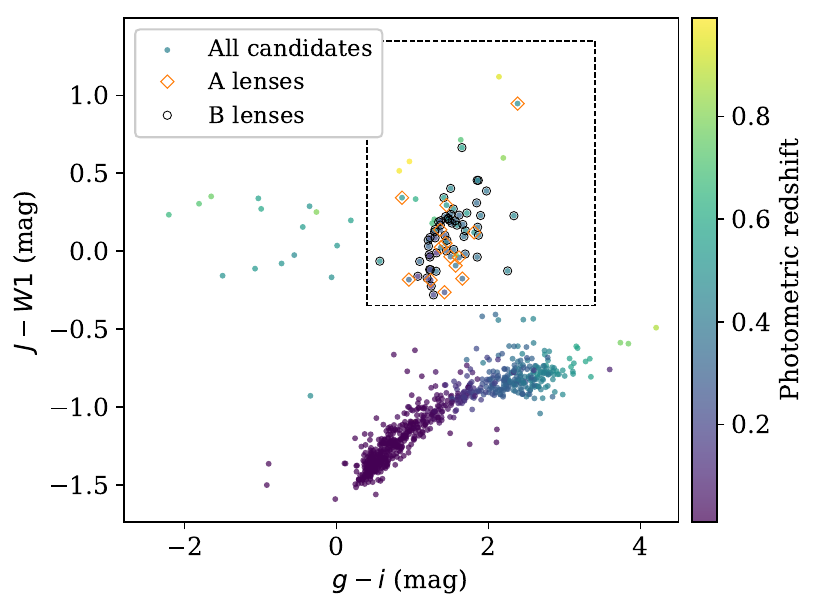}
    \caption{Color-color diagram used to distinguish stars and galaxies. The black dashed rectangle indicates our selection of lens candidates ($J-\mathrm{W1}>-0.35$, $g-i>0.4$), while stars are located in the $J-\mathrm{W1}\lesssim-0.5$ region (in AB system, extinction-corrected magnitudes). Photometric redshifts derived from SED analysis are shown as a color map, as additional redshift constraints were applied to exclude sources with unreliable photometric redshifts, particularly at $z<0.1$ and $z>0.6$. Best candidates (A lenses) were further selected based on the $\chi^2_r$ of the models and the proximity of the submillimeter sources.}
    \label{fig:J-W1}
\end{figure}

There are several procedures to separate stars from galaxies at optical and NIR wavelengths. For instance, following \cite{Fleuren2012}, color cuts can be employed using the SDSS and VIKING flux densities to define a star and galaxy locus on a color-color diagram. A similar approach was taken by \cite{Bakx2020} without SDSS data, selecting as galaxies the objects with colors $J-K_s>0$. This cut has been shown to include as many galaxies as possible without introducing too many stars, although similar results can be obtained using a $J-\mathrm{W1}$ color selection. For convenience, a threshold of $J-\mathrm{W1}>-0.35$ mag (in AB system) was adopted in this case, along with a $g-i>0.4$ cut to exclude ambiguous sources (see Fig. \ref{fig:J-W1}).

To further remove unreliable cases with poorly converged fits from the sample, we also excluded lens candidates with photometric redshifts outside the $0.1{-}0.6$ range, ending with a final sample of 68 candidate gravitational lenses from the initial 849. Additionally, optical surveys like the SDSS, VST and Pan-STARRS include a column in their catalogs that flags objects by type or class, based on photometric and point spread function analyses, which can be used as an independent check. In general, the two methods are consistent when such data are available.

In addition to stars, we searched for potential contamination from spiral galaxies, as they also have similar WISE colors to those of ellipticals and they would reduce the effectiveness of the lensing magnification. To investigate this, we examined the $u-r$ color distribution of the candidates alongside their stellar masses, since spiral galaxies are generally bluer. As expected, most lie above the green valley, a transition region between spirals and ellipticals \citep[e.g.,][]{Strateva2001, Baldry2004, Baldry2006, Schawinski2014}. Although this color-based distinction is not a strict classification criterion due to population overlap \citep{Baldry2004}, it serves as a statistical estimate of spiral contamination. Thus, we infer that our sample is predominantly composed of ellipticals, although some degree of spiral contamination cannot be entirely ruled out.

\subsection{Internal validation}\label{sec:internal_validation}
To assess how well galaxy properties are constrained through multi-wavelength SED fitting, we utilized the mock analysis module included in CIGALE. This module generates mock SEDs from the original best-fit models, providing a direct measure of how reliably the parameters can be recovered in our specific sample, as further discussed in Appendix \ref{sec:mock_analysis}.

On the other hand, for a robust identification of the most probable strong-lensed sources in our catalog, the reduced chi-squared value ($\chi^2_r$) of the best-fit models was used as a global quality estimator, combined with a constraint on small angular separations. Therefore, the final sample of 68 candidates was divided into two groups, A and B, ranking their strong lensing probability. In particular, only 15 sources are included in first group, all of which satisfy the following requirements: (i) having a sufficiently low reduced chi-squared value ($\chi^2_{r,\ \text{lens}} < 10$), and (ii) an angular separation of less than 5 arcsec from the H-ATLAS sources. The remaining 53 candidates are assigned to group B.

\begin{figure*}
\centering
\includegraphics[width=0.85\linewidth]{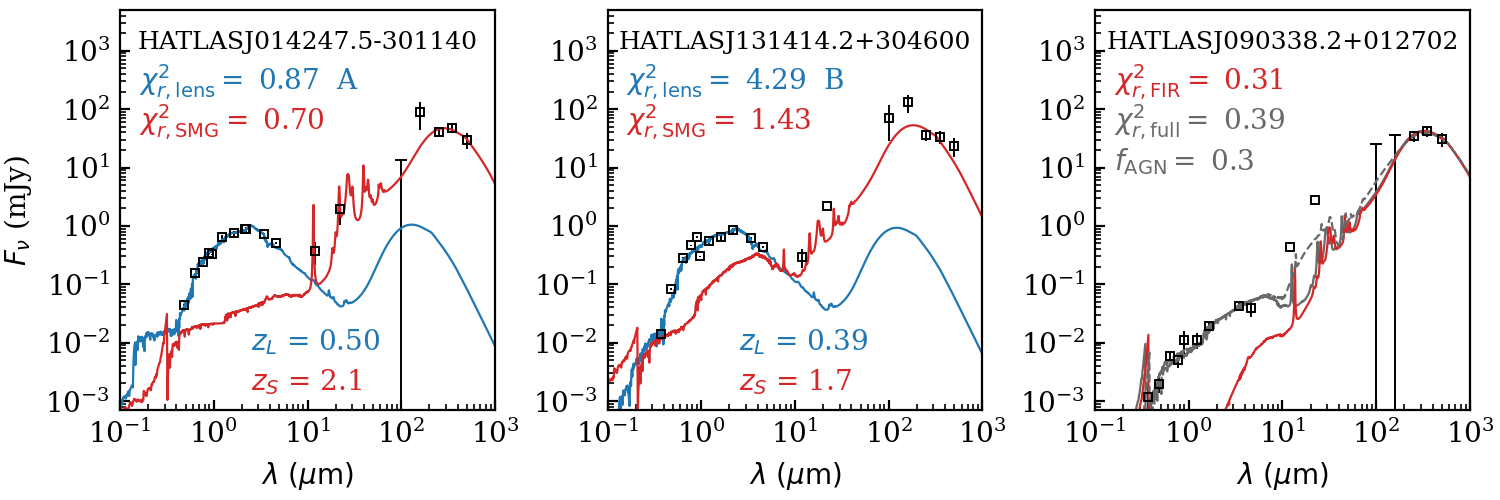}
\caption{Example of the best-fit models for a type A candidate (left), a type B candidate (center), and one of the unlensed SMGs selected. In the first two panels, the elliptical and SMG components are shown respectively in blue and red, together with their $\chi^2_r$ values and redshifts in matching colors. In the third panel, red indicates the fit to the far-IR component of the SED, while the fit including all available filters is shown in gray. In addition, the dashed gray curve represents an alternative fit including a modest AGN contribution. The complete set of best-fit models for the candidates can be found in Appendix \ref{sec:full_catalog}.}
\label{fig:particular cases}
\end{figure*}

The full set of best-fit models is included in Appendix \ref{sec:SED_models}, but a representative example of each type is displayed in Fig. \ref{fig:particular cases}. As we can see, observational data for A candidates are less scattered due to the $\chi_r^2$ constraint, while some candidates from group B exhibit small discontinuities in their SEDs, which could be related to blending with nearby optical counterparts. Moreover, some noticeable excesses can sometimes be seen in PACS data, which are mostly upper limits due to its reduced sensitivity. However, when such excesses fall outside the uncertainty range, they could indicate second peak of dust emission from another source at a lower redshift, potentially a spiral galaxy or a nearby dusty pre- or post- main sequence star. But given the lower resolution of PACS, this emission may not be linked to the lens but could be from another source in the field. Calibration issues or data errors could also explain these excesses.

\subsection{Single and multiple counterparts}
Along with stellar contamination and SED fitting uncertainties, source multiplicity and blending are some of the main concerns of this analysis. To mitigate this issue, one possibility is to limit our analysis to isolated sources, i.e., sources with no other counterpart in any of the catalogs used. However, this approach would reduce the number of lenses in our sample to three (or up to eight if we exclude the WISE isolation condition, but maintaining it in optical and NIR data). Table \ref{tab:list_candidates} indicates which sources are isolated in each wavelength range (optical, near-, and mid-IR). Since their properties are not very different from those of the other candidates, we can conclude that blending effects do not introduce any strong statistical bias in those candidates with more than one counterpart.

\subsection{External validation}\label{sec:external_validation}
As mentioned in the first section, several catalogs of strongly lensed objects have been compiled based on their high submillimeter flux densities at 500 $\mu$m or their association with VIKING counterparts (\citealt{Nayyeri2016,Negrello2017,Bakx2020,Bakx2024}; some of which have been confirmed through high-resolution imaging), or derived from statistical analyses based on different observables \citep[][]{GON12,GON19}. But, interestingly, none of the sources in our sample match any of the confirmed lens catalogs, which means that this is a completely new and independent method for searching gravitational lensing candidates.

As shown in Fig. \ref{fig:EllipticalsW12 vs W23}, all the WISE counterparts of the lensed SMGs included in these catalogs fall outside our selection of elliptical galaxies, since their colors more closely resemble those of starburst galaxies. This suggests that these lenses are considerably fainter than those in our sample, allowing the SMG emission to dominate at WISE wavelengths. Moreover, some of these lensed SMGs could be intrinsically brighter as a consequence of the flux density selection of these methods. As a result, although our approach does not impose any flux density threshold, it is inherently biased toward systems with brighter foreground lenses.

\subsection{Final dataset}
After the selection process described above, a total of 68 new gravitational lens candidates were identified. Of these, 15 were selected as the most probable strong lensing events based on the proximity of the foreground lenses and the quality of the fitting (Sect. \ref{sec:internal_validation}). The resulting catalog of strong and weak lensing candidates is presented in Appendix \ref{sec:full_catalog}, together with some of their most relevant physical parameters derived from SED analysis, including photometric redshifts, the reduced chi-squared values of the best-fit models, stellar masses, an isolation flag for optical and IR catalogs, and a lensing rank assigned to each lens. It is also worth noting that, due to our conservative approach aimed at minimizing the inclusion of stars and unreliable cases, additional gravitational lenses may exist outside our sample. However, relaxing the selection criteria would increase the risk of false positives, requiring additional effort to exclude them and compromising the overall reliability of the sample.

\subsection{Unlensed SMGs}\label{sec:non-lensed_validation}
As discussed before, we used a small sample of unlensed SMGs to test our analysis and identify potential issues affecting our results. To ensure the robustness of this method, it is essential to generate a reliable test sample that not suffers from the main limitations of the lensed dataset, such as the limited spectral coverage of each galaxy, positional uncertainties and source multiplicity during cross-matches, and the high contamination from stellar objects. While, in principle, the selected sample of unlensed SMGs is less likely to be affected by these issues, given their location in the WISE color-color diagram (Fig. \ref{fig:EllipticalsW12 vs W23}), the following selection criteria were applied to minimize the risk of contamination by unreliable data: (i) having an angular separation within 10 arcsec from the parent Herschel sources, (ii) a maximum reduced chi-square value of $\chi^2_r = 1$ to ensure a robust fit, (iii) a maximum photometric redshift of $z=4$ in the full SED fit to guarantee good convergence, and (iv) a nonstellar SDSS classification. In addition, we also required these sources have data in optical or NIR wavelengths, ending with a final sample of 250 objects.

\section{Discussion}\label{sec:discussion}
\begin{figure*}
    \centering
    \includegraphics[width=0.65\linewidth]{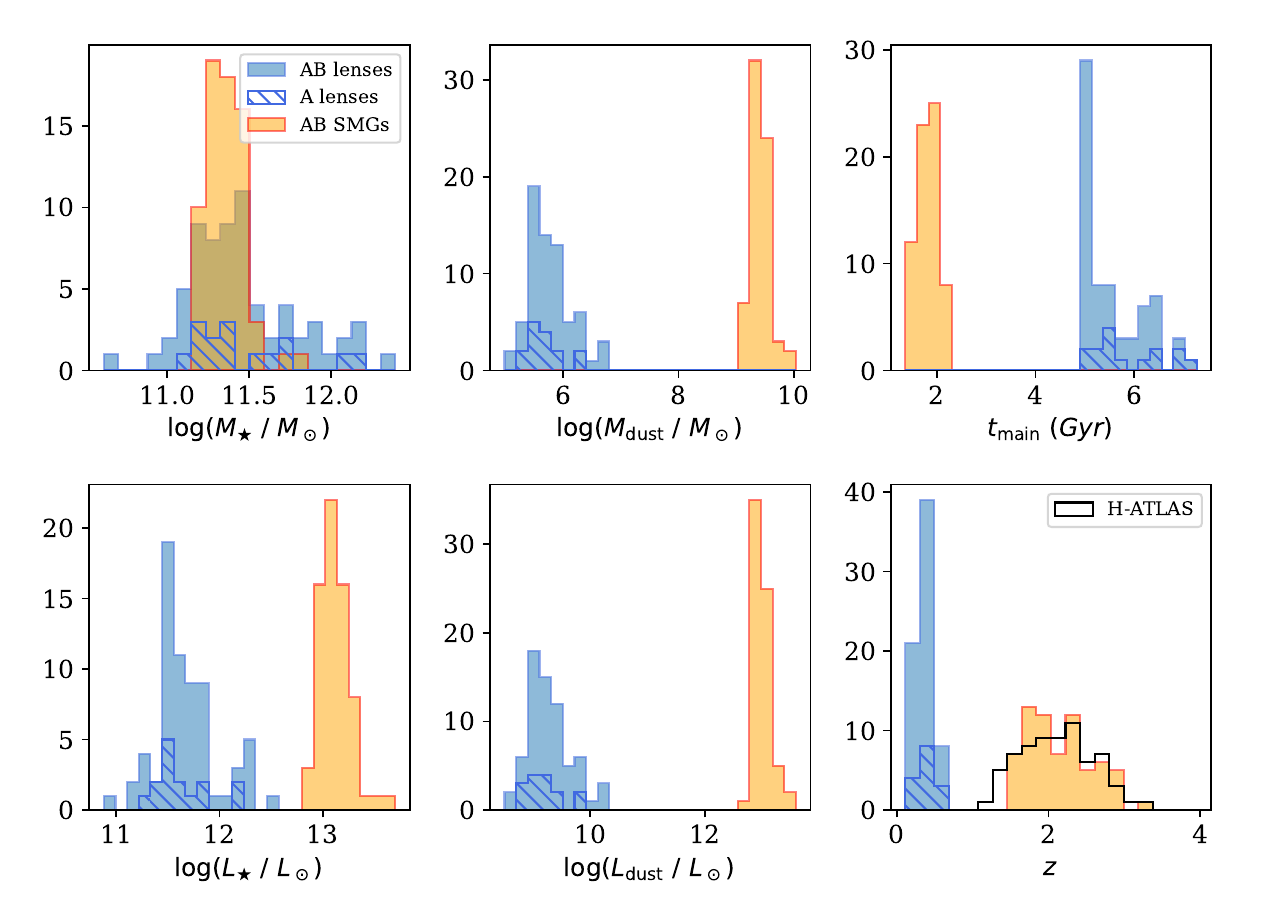}
    \caption{Main physical parameters of the lenses (filled blue) and lensed submillimeter (filled orange) candidates, as retrieved by the Bayesian analysis of the SED fitted models by CIGALE. The striped blue histograms represent the properties of the selected 15 best candidates from the lens sample. In the last panel, photometric redshifts from \cite{GON19}, obtained with the fitting by the SMM J2135-0102 template to the dust emission peak, are shown in black for comparison purposes.}
    \label{fig:stat}
\end{figure*}

\begin{figure}
\centering
\includegraphics[width=0.9\linewidth]{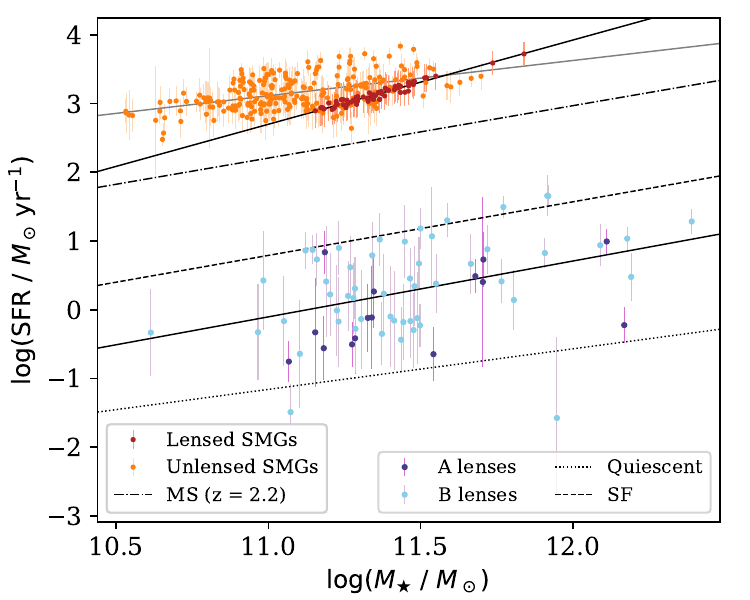}
\caption{SFR–$M_\star$ relation for the lens galaxies and SMGs in our sample. For comparison, the sample of unlensed SMGs is also shown, along with the quiescent (dashed line) and star-forming (dotted line) evolutionary tracks for early-type galaxies from \citet{Paspaliaris2023}, and the star-forming main sequence (MS) at $z = 2.2$ from \citet{Speagle2014} (dash-dotted line). Moreover, the solid lines represent the fits to our data.}
\label{fig:SFR_Mstar}
\end{figure}

\begin{figure*}
    \centering
    \includegraphics[width=0.65\textwidth]{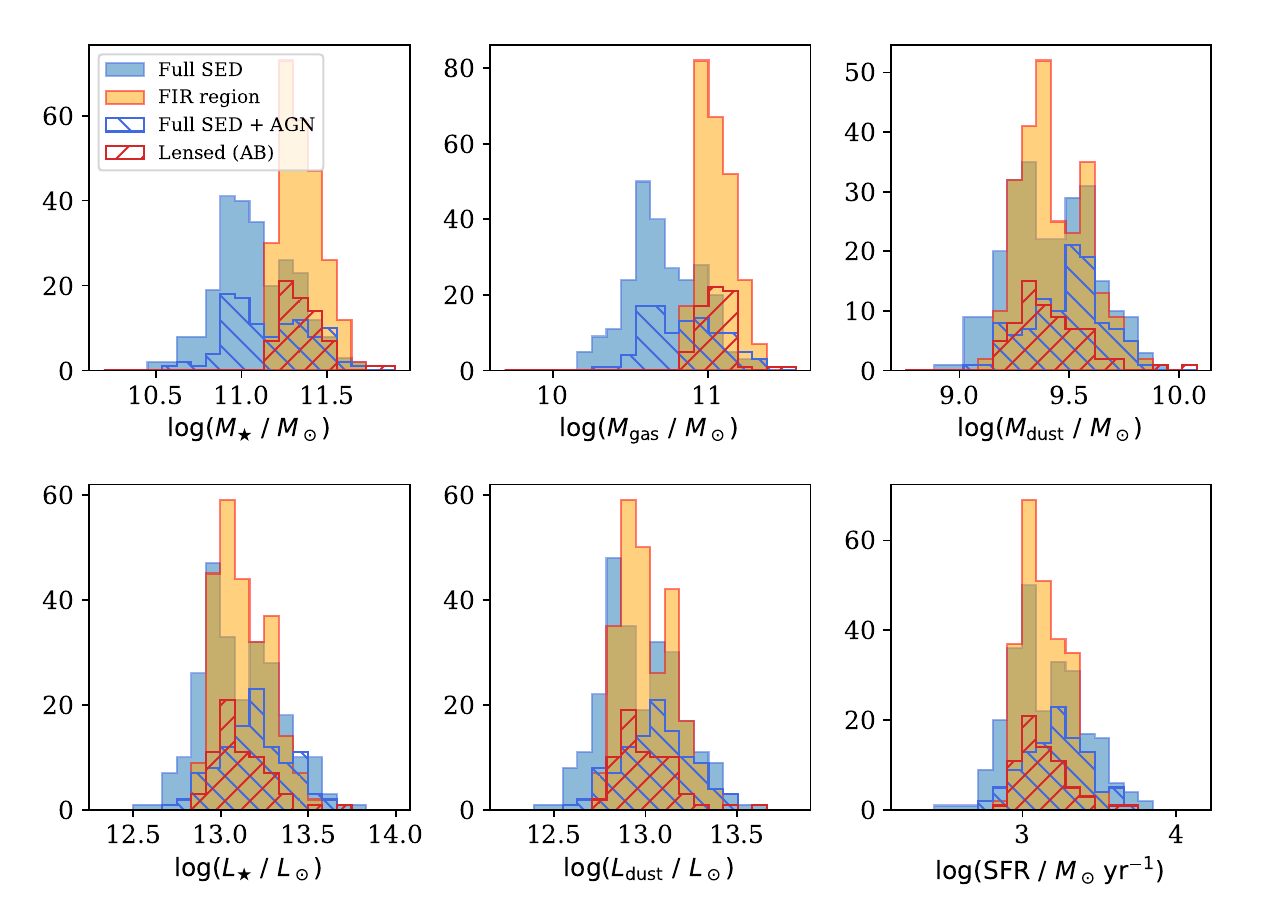}
    \caption{Comparison of the obtained properties of the unlensed SMG sample using the whole set of filters available (shaded blue), and the ones obtained by restricting to FIR data (shaded orange). Lensed SMG candidates are shown in red, while the striped blue histogram represents the unlensed sources with a small AGN component.}\label{fig:starburst_stat}
\end{figure*}

\subsection{General properties of the candidate lenses and lensed SMGs}\label{sec:discussion_SMGs}
The parameters obtained from SED analysis provide valuable insights into the nature and properties of our lenses and lensed candidates. A selection of the main derived quantities (stellar and dust masses and luminosities, age of the main stellar population, photometric redshift, and SFR) is presented in Figs. \ref{fig:stat} and \ref{fig:SFR_Mstar}.

First, marked differences are found between ellipticals and SMGs: the latter are considerably younger and exhibit significantly higher redshifts, SFRs, dust masses and luminosities, whereas ellipticals show the opposite trend. In general, these values align well with previous studies, although there are some exceptions.

For instance, the obtained SFRs for the SMGs, ranging from $(0.8{-}5)\times 10^3\ M_\odot\, \text{yr}^{-1}$, are in line with previous works \citep[][and references therein]{Casey2014}, as well as stellar masses, between $(1{-}7)\times 10^{11}\ M_\odot$, with a median value of $2\times 10^{11}\ M_\odot$ \citep{Borys2005,Dudzeviciute2020}. However, the typical stellar masses of the broad SMG population are still a subject of debate, with some studies suggesting lower values in the $10^{10}{-1}0^{11}\ M_\odot$ range \citep[e.g.,][]{Hainline2011,Negrello2014,daCunha2015}. These lower estimates are further supported by independent CO kinematic observations \citep{Engel2010}. The discrepancies likely arise from differences in the modeled SFH, IMF assumptions, or stellar population synthesis, as discussed by \cite{Michalowski2012} and \cite{Conroy2013}. In our case, the lack of optical and NIR data for the lensed galaxies prevents a precise measurement of the stellar parameters, making them more susceptible to strong biases introduced by model assumptions and data availability (see Sect. \ref{sec:comparison_SMGs}).

Stellar luminosities, with a median value of $1\times 10^{13}\ L_\odot$, are also consistent with the fraction of stellar emission absorbed by dust, which is ${\sim}77\%$ in nearly all cases, as expected from local ULIRGs \citep{Paspaliaris2021}. In contrast to stellar parameters, dust masses and luminosities are better constrained in submillimeter data, with median values of $\sim 2\times 10^9\ M_\odot$ and $\sim 9\times 10^{12}\ L_\odot$, respectively. However, it is important to note that the actual magnifications of these sources are unknown and likely contribute to boosting the observed luminosities and masses to some degree.

On the other hand, elliptical galaxies show much lower SFRs and dust content than SMGs, as expected given their older stellar population and quenching of the SFH. Their photometric redshift distribution, as well as their values of $L_\star$, $M_\text{dust}$, and $L_\text{dust}$, are also in good agreement with the literature. However, some objects exhibit quite extreme stellar masses ($>10^{12}\ M_\odot$), with a mean value of $2\times 10^{11}\ M_\odot$. This may result from inaccuracies in redshift estimations, since the most massive objects have systematically converged to higher redshifts, but also due to blending and source multiplicity in optical wavelengths, the degeneracy produced by the lack of data at UV and FIR wavelengths, or even the presence of a foreground galaxy cluster.

Figure \ref{fig:SFR_Mstar} shows the SFR-$M_\star$ relationship measured for the ellipticals and SMGs, which describes the typical behavior of star-forming galaxies at a given stellar mass and redshift, driven by gas accretion, internal feedback, and environmental effects. As we can see, the ellipticals in our sample appear scattered between the empirical main sequence and quiescent scenarios from \citet{Paspaliaris2023}, consistent with their more evolved stellar populations and lower star formation activity. In contrast, SMGs lie well above the main sequence at comparable redshifts, as expected for a population undergoing intense starbursts likely triggered by major mergers \citep[e.g.,][]{Daddi2007,Engel2010,Hainline2011,Magnelli2012}. Notably, these galaxies show a tighter correlation than ellipticals, as noted by previous studies (e.g., \citealp{Brinchmann2004,Elbaz2007,Wuyts2011,Magdis2012,Chang2015,Pearson2018}), and the slope observed for the unlensed population is a bit flatter \citep[e.g.,][]{Paspaliaris2021}.

\subsection{Comparison of the lensed candidates with the unlensed sample}\label{sec:comparison_SMGs}
For the unlensed dataset, we repeated the analysis applied to the lensed candidates, using the same SED models and input parameters as in the previous case. However, in this case the fitting was performed twice: first using the full set of available filters, from UV to submillimeter, and then restricting the fit to the FIR region only. This dual approach allowed us to evaluate potential biases introduced when the fitting is performed in cases where shorter wavelength data is not available, as was the case for the lensed SMGs.

The results of this analysis revealed a considerable discrepancy in $M_\star$ and $M_\text{gas}$ estimations (Fig. \ref{fig:starburst_stat}), driven by the degeneracy between the stellar mass and dust attenuation. Indeed, the expectation value for dust extinction (not included in the figure) converged to a value of $E(B-V)\simeq 1.3$ in all cases, since all of the input values were equally favored in the fitting of FIR data. This degeneracy is broken when considering the full SED of the galaxies, as it provides a direct measure of the energy balance between photons absorbed in the UV/optical and re-radiated in the IR, thereby allowing better constraints on dust attenuation and stellar parameters. Extrapolating this measured bias to the lensed sources, we expect our SMGs to be about $2{-}3$ times less massive, leading to a median value of $M_\star\sim 7\times 10^{10}\ M_\odot$. This aligns more closely with lower-mass predictions of \cite{Engel2010, Hainline2011, Negrello2014, daCunha2015}.

However, recall that this sample of unlensed galaxies may not be a fully representative subset of the entire population of SMGs, as we are selecting only those with available optical or NIR data. This results in a strong observational bias toward intrinsically bright galaxies that exceed the survey's detection limits, implying lower dust attenuation and thus favoring the detection of less obscured galaxies at high redshifts. Nevertheless, this does not seem to affect $L_\star$, as the estimated values are practically the same in the two fits (Fig. \ref{fig:starburst_stat}). This is because stellar luminosity is computed from the intrinsic (i.e., unattenuated) spectra of the galaxy's stellar population, as determined by the assumed IMF, SFH, and stellar population synthesis models, so it is less dependent on dust attenuation.

Conversely, FIR-based fits excel at determining SFRs and dust properties, and they are indeed well recovered with respect to the full SED fits (Fig. \ref{fig:starburst_stat}). These values are also very similar to those from the lensed dataset. In view of these results, we can conclude that the parameters obtained for the lensed candidates at FIR wavelengths are robust and reflect the true properties of these objects, with the exception of $M_\star$, $M_\text{gas}$, and $E(B-V)$, which may be significantly biased.

In addition, as shown in Fig. \ref{fig:starburst_stat}, the inclusion of an AGN component does not introduce any significant bias in the estimated properties, although the AGN fraction in the models is small (up to 30\%, as none of the 50\% models were used). These results are consistent with previous works \citep[e.g.,][]{Michalowski2014} and can be explained by the fact that most of the emission from an obscured AGN occurs in the MIR portion of the SED, having little effects in other regions.

\begin{figure}
    \centering
    \includegraphics[width=0.8\linewidth]{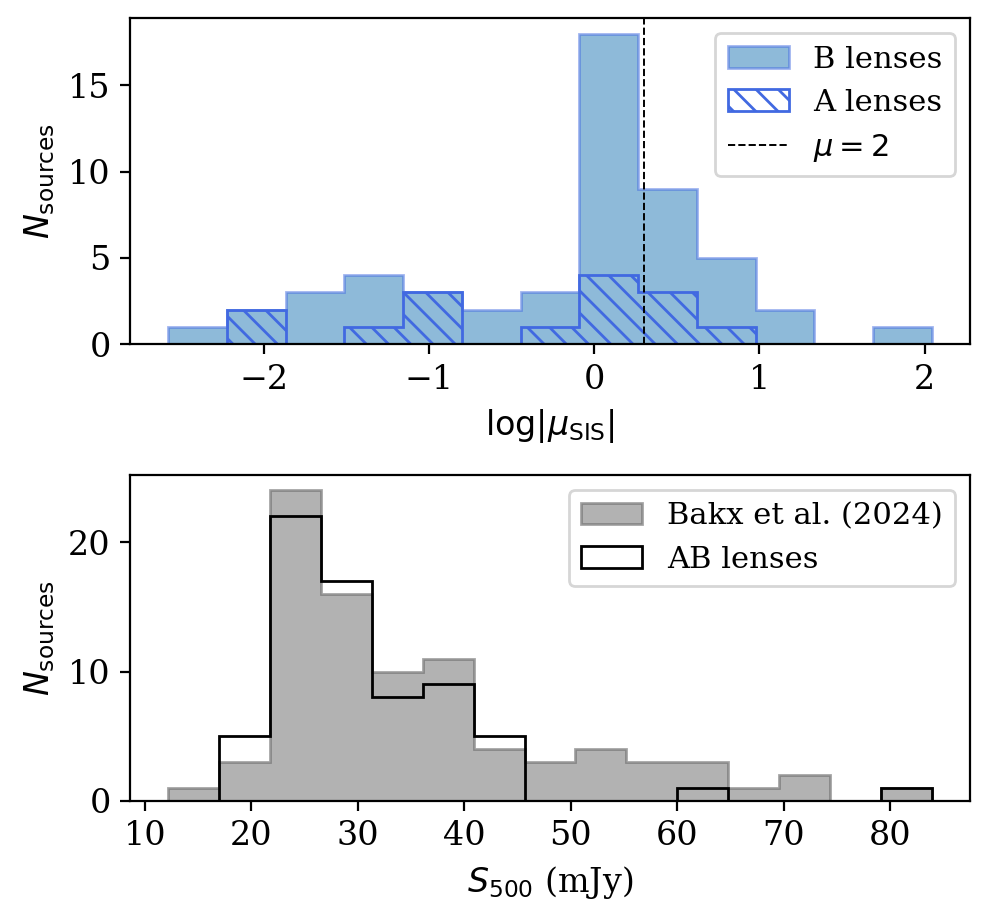}
    \caption{Estimated magnification of the lensed candidates (top panel), for the A and B type lenses, assuming a SIS halo model and using the masses and redshift of the foreground and background galaxies in the calculation. The value of $\mu=2$ is indicated for comparison purposes as a dashed black line. Despite the large uncertainties of this approach, it appears that most candidates have values of $\mu\gtrsim 1$ compatible with the weak magnification regime, while higher magnification and demagnification values would require a more accurate modeling. The bottom panel shows the distribution for the flux densities at 500 $\mu$m.}
    \label{fig:magnification_AB}
\end{figure}

\subsection{Estimation of the lensing magnification}
By utilizing some of the physical properties calculated before, we were able to estimate other quantities more specifically related to the lensing phenomena, such as the Einstein radii of the lenses and the expected magnification of the background sources. However, these calculations should be taken with caution due to the large uncertainties inherent to SED analysis and the intrinsic degeneracies and systematic errors, which tend to amplify when propagated through the calculations. Estimated uncertainties are usually $\sim2{-}5$ times larger than the actual magnification values. Despite these important limitations, our estimates can still be useful in a statistical sense, providing a rough estimation of the magnification of the lens candidates and offering an alternative way to test the reliability of the identified lenses.

Using the stellar masses and photometric redshifts of the lenses, we applied the redshift-dependent parametrization of the stellar-to-halo mass relation from \cite{Moster2013} to compute the expected mass of the dark matter halos surrounding these galaxies. This allowed us to estimate the total mass of each lens as the sum of the masses from the stellar, gas, dust, and dark matter halo components, i.e., $M_\text{lens}=M_\star+M_\text{gas}+M_\text{dust}+M_h$. Subsequently, cosmological distances were computed with the Python version of CosmoCalc\footnote{For an online implementation, see \url{https://www.astro.ucla.edu/~wright/CosmoCalc.html}.} \citep{CosmoCalc}, using the photometric redshifts and cosmological parameters from \cite{Planck2020}.

As outlined in Appendix \ref{sec:Theoretical_framework}, these quantities can be used to estimate the magnification of the lensed candidates, assuming a given halo mass distribution. For simplicity, we adopted a singular isothermal sphere (SIS) model. Despite its idealized nature, with an infinite size and central divergence, it fits reasonably well within our purposes. Differences with more complex models, required for precise measurements, would be greatly outnumbered by our large uncertainties, so they are not expected to introduce significant differences in this case. The resulting magnifications are shown in the top panel of Fig. \ref{fig:magnification_AB}, where the logarithm of their absolute values is plotted for improved readability. As we can see, most candidates are compatible with the weak lensing magnification scenario ($\mu\simeq 1{-}1.5$), even though some cases are well above this value, while others suggest strong demagnification effects ($|\mu|<1$). However, these extreme magnification or demagnification values are associated with extremely large values of the Einstein radii ($\theta_E>100$ arcsec) and lens masses ($M_{\text{lens}} \gg 10^{14}\ M_\odot$), so they would require better data and a more careful modeling of their SEDs. This is because at high masses, the stellar-to-halo mass relation becomes highly uncertain and can lead to divergent results. Nevertheless, this does not necessarily rule out the lensing nature of these candidates, and they are indeed interesting targets for future observations and more detailed analyses. These high lensing masses and Einstein radii could also be explained by foreground galaxy clustering.

In fact, our method opens the possibility of identifying and studying low-magnification lensing events on a case-by-case basis. These are typically overlooked in follow-up studies, which tend to focus on strongly lensed sources due to their higher magnification and easier confirmation, while low-amplification events have so far remained accessible only through statistical weak lensing analyses. Their flux density distribution at 500 $\mu$m, shown in the bottom panel of Fig. \ref{fig:magnification_AB}, is similar to that from \cite{Bakx2024} and consistent with the average SMG population in H-ATLAS, but considerably lower than in earlier works focused on strongly lensed sources.

\section{Conclusions}\label{sec:conclusions}
In this study, we propose a new and independent methodology to identify gravitational lens candidates within the H-ATLAS and AllWISE surveys. In contrast to previous methods, this catalog is not biased toward bright submillimeter strongly lensed galaxies, allowing for the detection of fainter, lower-magnification lensing events that may be missed by traditional flux-limited selections. This, in turn, provides the opportunity to identify and study such cases individually, which could provide complementary insights beyond statistical weak lensing analyses. However, this comes at the cost of reduced efficiency in identifying truly lensed systems, requiring a more rigorous filtering process to discard unreliable or ambiguous candidates.

As a practical application, we focused on high-redshift SMGs ($1.2<z<4.0$) from H-ATLAS whose mid-IR emission deviates from that expected for starburst galaxies and may instead originate from foreground ellipticals, potentially acting as gravitational lenses. These candidates were identified by searching for WISE counterparts within 18 arcsec and selecting those with WISE colors in the range $0.5 < \mathrm{W2} - \mathrm{W3} < 1.5$ mag, consistent with the expected colors of elliptical galaxies. This method assumes that the mid-IR emission from the lenses dominates over that of their companion SMGs, so they can be observed by WISE. As a result, it may be biased toward brighter foreground lenses compared to previous selection techniques. Furthermore, since WISE colors of elliptical galaxies partially overlap with those of stars, stellar interlopers were further removed based on a $J-\mathrm{W1}$ color cut and photometric redshift criteria. After excluding such unreliable cases, the final sample comprises 68 new gravitational lens candidates.

To inspect the lensing nature of these candidates and estimate their rough amplifications, we performed SED fitting analysis with CIGALE from the UV to the submillimeter. Despite the inherent uncertainties of this approach, it allowed us to constrain key physical parameters such as stellar and dust masses and luminosities, SFRs, or photometric redshifts for both lenses and background SMGs. In addition, the estimated magnifications of the lensed candidates are generally low, compatible with the weak lensing scenario ($\mu\simeq 1{-}1.5$), though some sources exhibit extreme values of magnification or apparent demagnification. These outliers would require more detailed SED modeling to obtain reliable estimates of their physical properties. Nevertheless, this does not necessarily rule out the lensing nature of these candidates, and they remain promising candidates for follow-up observations.

Finally, it is worth noting that our conservative approach, aimed to minimize stellar contamination and exclude uncertain cases, may have led to the omission of additional gravitational lenses beyond our selection. Future efforts could refine this methodology to recover potential candidates, and high-resolution follow-up observations will be essential to confirm the lensing nature of these sources and to further investigate their physical properties.

\begin{acknowledgements}
JAC, JGN, LB, MMC, TJLCB, JMC, RFF and DC acknowledge the PID2021-125630NB-I00 project funded by MCIN/AEI/10.13039/501100011033/FEDER, UE. LB also acknowledges the CNS2022-135748 project funded by MCIN/AEI/10.13039/501100011033 and by the EU “NextGenerationEU/PRTR”. \\
The Herschel-ATLAS is a project with Herschel, which is an ESA space observatory with science instruments provided by European-led Principal Investigator consortia and with important participation from NASA. The H-ATLAS web-site is http://www.h-atlas.org. WISE is a joint project of the University of California, Los Angeles, and the Jet Propulsion Laboratory/California Institute of Technology, funded by NASA. Complementary imaging of the H-ATLAS regions is being obtained by a number of independent survey programs including GALEX MIS, VST KIDS, VISTA VIKING, WISE, Herschel-ATLAS, GMRT and ASKAP providing UV to radio coverage.\\
This research has made use of the Python packages \texttt{astropy} \citep[][and references therein]{astropy2022}, \texttt{ipython} \citep{ipython}, \texttt{matplotlib} \citep{matplotlib}, \texttt{numpy} \citep{numpy}, \texttt{scipy} \citep{scipy}, and \texttt{dustmaps} \citep{dustmaps}.
\end{acknowledgements}

\bibliographystyle{aa}
\bibliography{myfile}

\begin{appendix}
\section{Mock analysis}\label{sec:mock_analysis}
\begin{figure}[ht!]
    \centering
    \includegraphics[width=\linewidth]{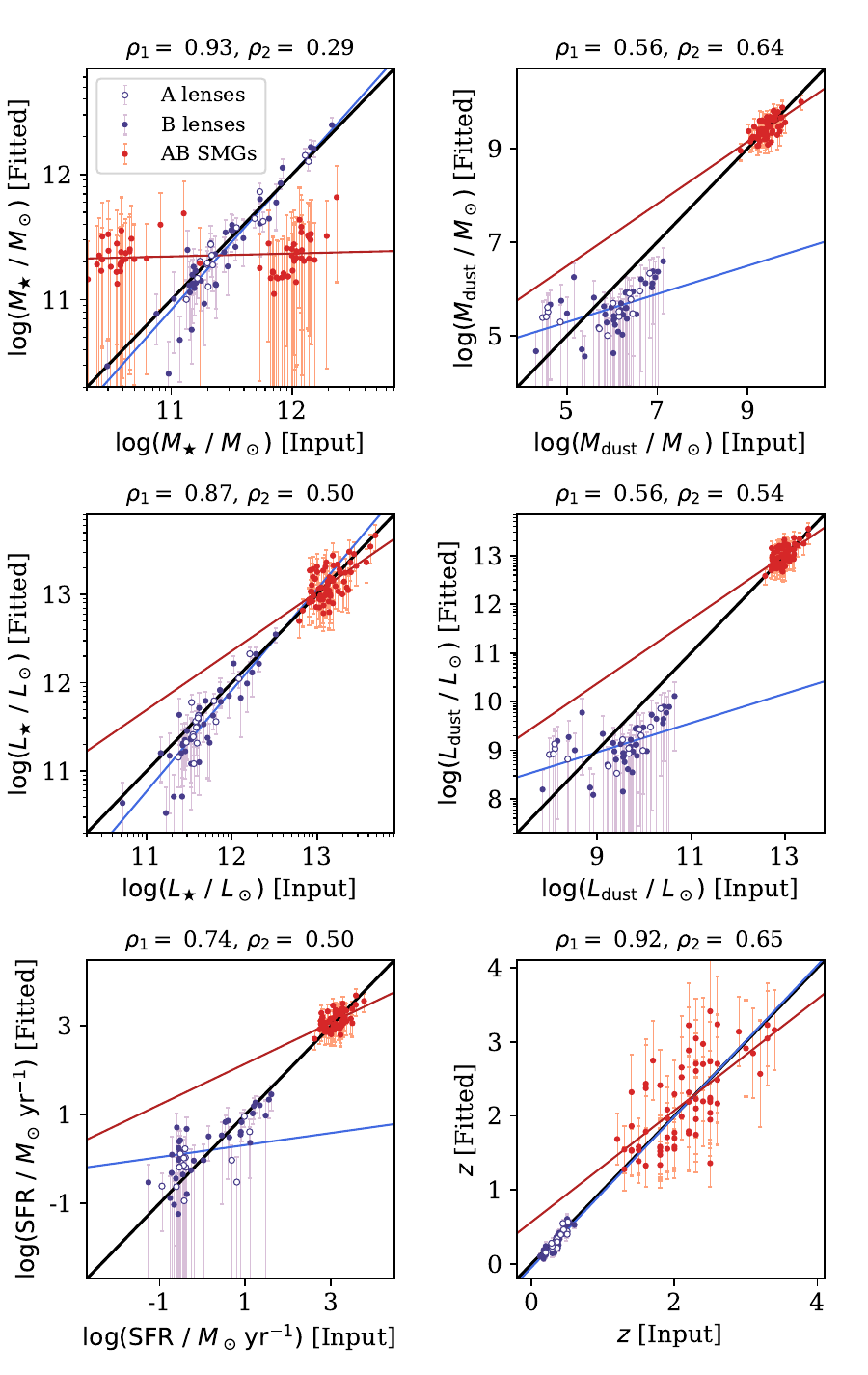}
    \caption{Comparison between the best-fit parameters (input values, plotted on $x$-axis) and the mock parameters (fitted values, plotted on $y$-axis) estimated with CIGALE for the lens candidates (blue) and the lensed SMGs (red). The solid black line represents the one-to-one relation, while the blue and red solid lines are the linear regressions for the lenses and SMGs. At the top of each panel, the calculated value of the Spearman's coefficient ($\rho$) is shown for both datasets.}
    \label{fig:mock_analysis}
\end{figure}

\begin{figure}[ht!]
    \centering
    \includegraphics[width=\linewidth]{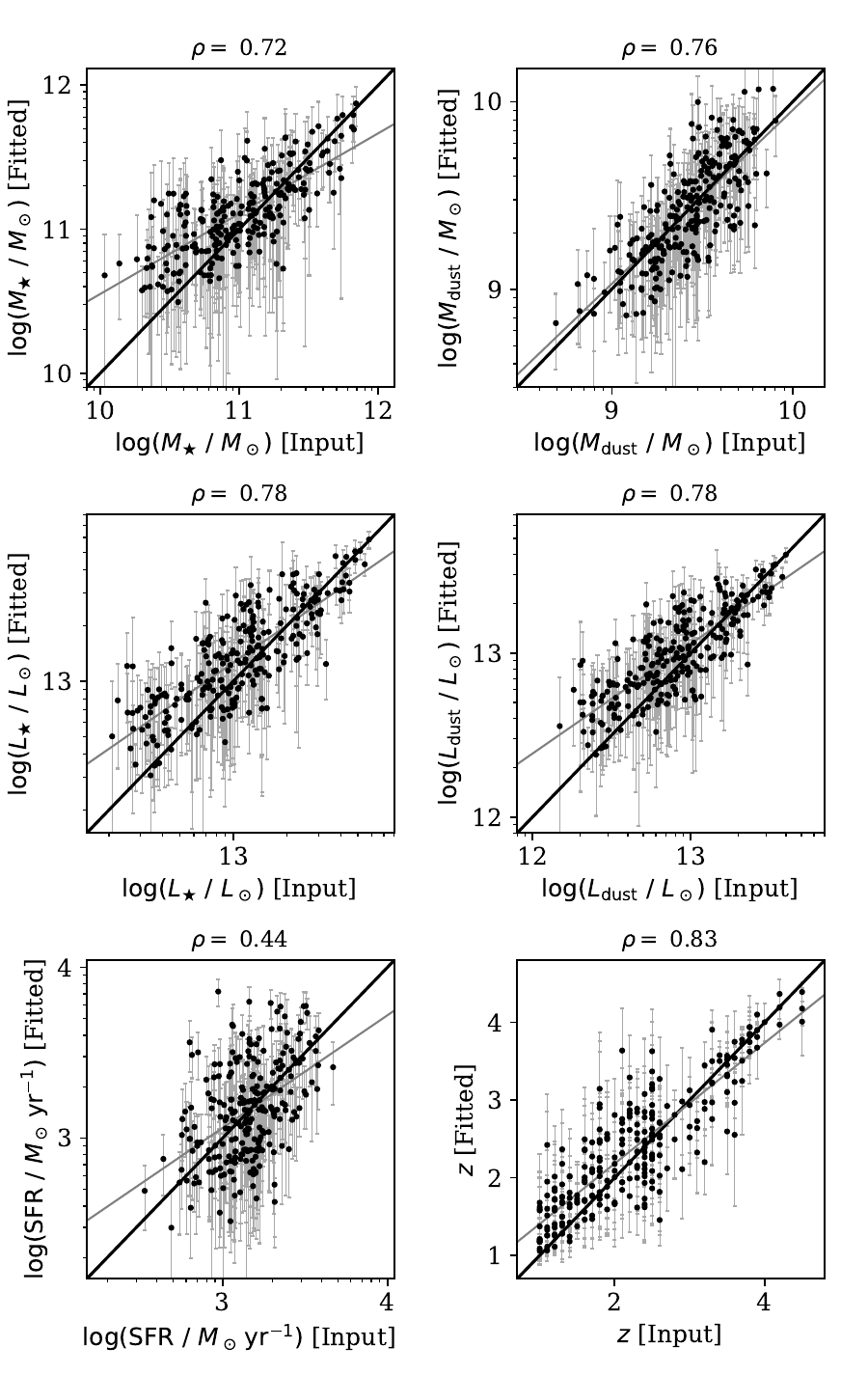}
    \caption{Comparison between the best-fit parameters and the mock parameters estimated with CIGALE for the unlensed sample of SMGs. As in Fig. \ref{fig:mock_analysis}, the solid black line represents the one-to-one relation, and the gray solid line is the linear regression for the entire dataset.}
    \label{fig:starburst_mock_analysis}
\end{figure}

In order to examine the accuracy and precision of the parameters obtained from the multi-wavelength SED fitting analysis, we made use of the mock analysis module included in CIGALE. This module generates a mock SED for each galaxy based on the best-fit model, allowing the fluxes to vary within the uncertainties by a random amount taken from a Gaussian distribution with the same standard deviation as the observations. This mock SED is then re-fitted, so the retrieved parameters from the mocks can be compared with their exact values from the input models. This provides us with a direct measure of the accuracy of the SED fitting for the specific sample we are using.

The results of the mock analysis performed for our gravitational lensing candidates are shown in Fig. \ref{fig:mock_analysis}, separately for the ellipticals and SMGs. The input values of the best-fit model are represented on the $x$-axis, compared to the obtained mock values on the $y$-axis. In addition, the calculated value of the Spearman's coefficient ($\rho$) is shown at the top of each panel for both datasets, although a minimum of $\sim$500 sources will be needed for a robust calculation. However, it serves as a quantitative estimation of how well the properties of these mock galaxies are recovered.

In general, most of the stellar parameters of the lenses and dust properties of the SMGs show a good correlation with the input values, but there are noticeable biases in the SFR and dust properties of the lenses and the stellar and gas masses of the SMGs. These parameters are more weakly constrained for several reasons, but in case of the SMGs, this is mainly due to the lack of data at UV/optical wavelengths, which introduces a strong degeneracy between dust attenuation and stellar mass, making it impossible to constrain the energy balance between the radiation absorbed in the UV/optical and that re-emitted in the IR. As a result, the best-fit models show a strong bimodality, which is averaged out in the Bayesian estimation. These effects are smaller for the dust properties of the lenses, as dust attenuation can be more effectively constrained by the modulation of the galaxy's stellar emission in the UV and optical SED.

Moreover, there is a significant scatter in the redshift estimation of the SMGs, but this is also a consequence of the small number of photometric bands available. At submillimeter wavelengths, redshift calculations are based on the position of the dust emission peak, which in our case is mainly described with the three SPIRE bands, so any small variation in these values may cause a large difference in the redshift estimation (but at least maintaining always the high-redshift nature of the SMGs with $z>1$).

In addition, Fig. \ref{fig:starburst_mock_analysis} shows the mock analysis performed for the testing subset of unlensed SMGs. Compared to our lens sample, the overall performance of the SED fitting and analysis is quite good, despite the large scatter and uncertainties. However, SFR and redshift estimations remain somewhat imprecise in this case, though they improve significantly when fitting only to the submillimeter data.

\section{Theoretical framework}\label{sec:Theoretical_framework}
The magnification ($\mu$) produced by a gravitational lens on a background source depends on the mass distribution of the lens, the geometry of the system, and their relative alignment respect to the observer, including their angular diameter distances. Therefore, in order to estimate the resulting magnification of that system, it is necessary to know the mass density profile of the lens halo. There are several models available for that purpose, like the Navarro-Frenk-White profile \citep[NFW;][]{NFW1996}, power-law distributions like the $R^{1/n}$ Sérsic profile \citep{Sersic1963, Sersic1968}, or composite models \citep[e.g.,][]{Lapi2012}, but one of the simplest models is the singular isothermal sphere (SIS), which describes an ideal, spherically symmetric distribution of self-gravitating particles whose velocities follow a Maxwell-Boltzmann distribution. The density profile of such systems is given by
\begin{equation}
\rho_\text{SIS}(r)=\frac{\sigma_v^2}{2\pi Gr^2},
\end{equation}
where $\sigma_v$ is the unidimensional velocity dispersion of the particles in the halo, which is approximately $\sigma_v^2\simeq V_h/\sqrt{2}$, with a Keplerian velocity of rotation given by $V_h=GM_h/r_h$, and $M_h$ and $r_h$ are the mass and radius of the halo. For that density profile, we can compute the Einstein radius of the system,
\begin{equation}
\theta_E=\frac{4\pi\sigma_v^2}{c^2}\frac{D_{LS}}{D_S}=\frac{2\pi GM_h}{c^2r_h}\frac{D_{LS}}{D_S},
\end{equation}
where $D_{LS}$ and $D_S$ are the angular diameter distances of the lens to the source and the source to the observer. We note that when using angular diameter distances, the lensing equation remains valid even in curved spacetimes, although in general the condition $D_{LS} = D_S-D_L$ is not necessarily true.

For all practical purposes, angular distances can be computed from the measured redshifts of the lens and the lensed galaxies, and the total mass of the halo can be approximated by the mass contained inside the virial radius at which the density of the halo is 200 times the critical density, 
\begin{equation}
r_{200} = \left(\frac{3M_{200}}{4\pi 200\rho_c}\right)^{1/3},
\end{equation}
with
\begin{equation}
\rho_{crit}(z)=\frac{3H^2(z)}{8\pi G}
\end{equation}

In addition, if the angular position of the source lies within the Einstein radius ($\beta < \theta_E$), it can be easily shown that two images are produced at $\theta_\pm = \beta \pm \theta_E$, with magnification
\begin{equation}
\mu_{SIS}=\frac{|\theta|/\theta_E}{|\theta|/\theta_E-1}
\end{equation}
Since $\theta_+ > \theta_E$, its magnification is always greater than 1. In contrast, the other image can undergo strong demagnification as $\theta_-$ approaches to zero.

On the other hand, if the source lies outside the Einstein radius, only a single image will be observed at a distance $\theta = \theta_+ = \beta + \theta_E$.

\section{Best-fit SED models}\label{sec:SED_models}
The SEDs of the 68 lens and lensed candidates included in this catalog are presented here (Fig. \ref{fig:SED_fits_1}). The best-fit SED models of each galaxy are plotted, along with the reduced chi-square values, photometric redshifts, and lens type (A or B). A more complete set of parameters is specified on Table \ref{tab:list_candidates} for each candidate.

\begin{figure*}
\centering
\includegraphics[width=0.8\textwidth]{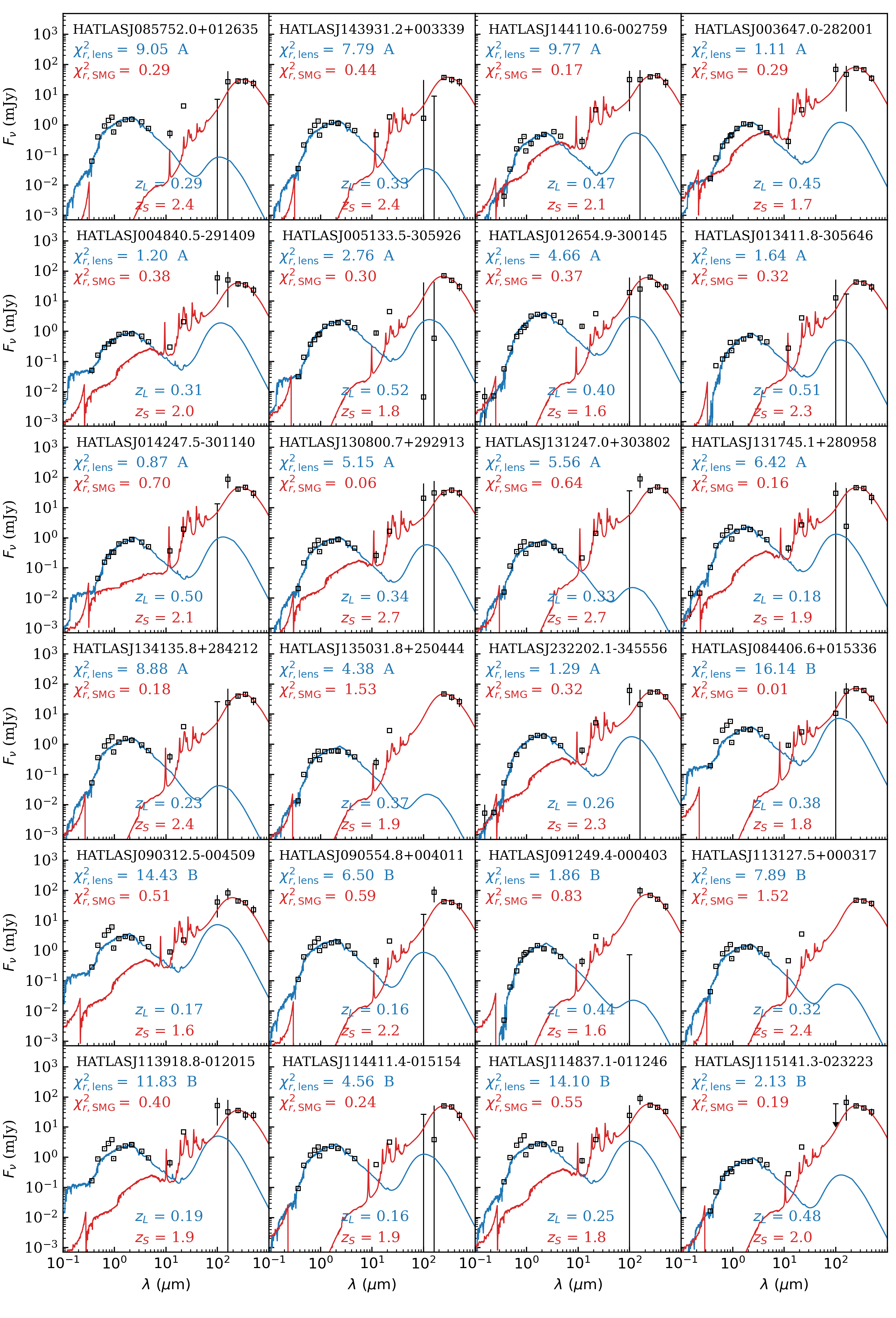}
\caption{Comprehensive collection of the best-fit models generated by CIGALE for the 68 lensing candidates in our catalog (in blue for the lens component; in red for the lensed SMGs). The observations for each galaxy, along with their uncertainties, are indicated by black open squares. In addition, for each component of the SEDs, the reduced chi-square values, photometric redshifts, and lens classification (type A or type B) are provided, following the same color-coding scheme.}\label{fig:SED_fits_1}
\end{figure*}

\begin{figure*}
\ContinuedFloat
\centering
\includegraphics[width=0.8\textwidth]{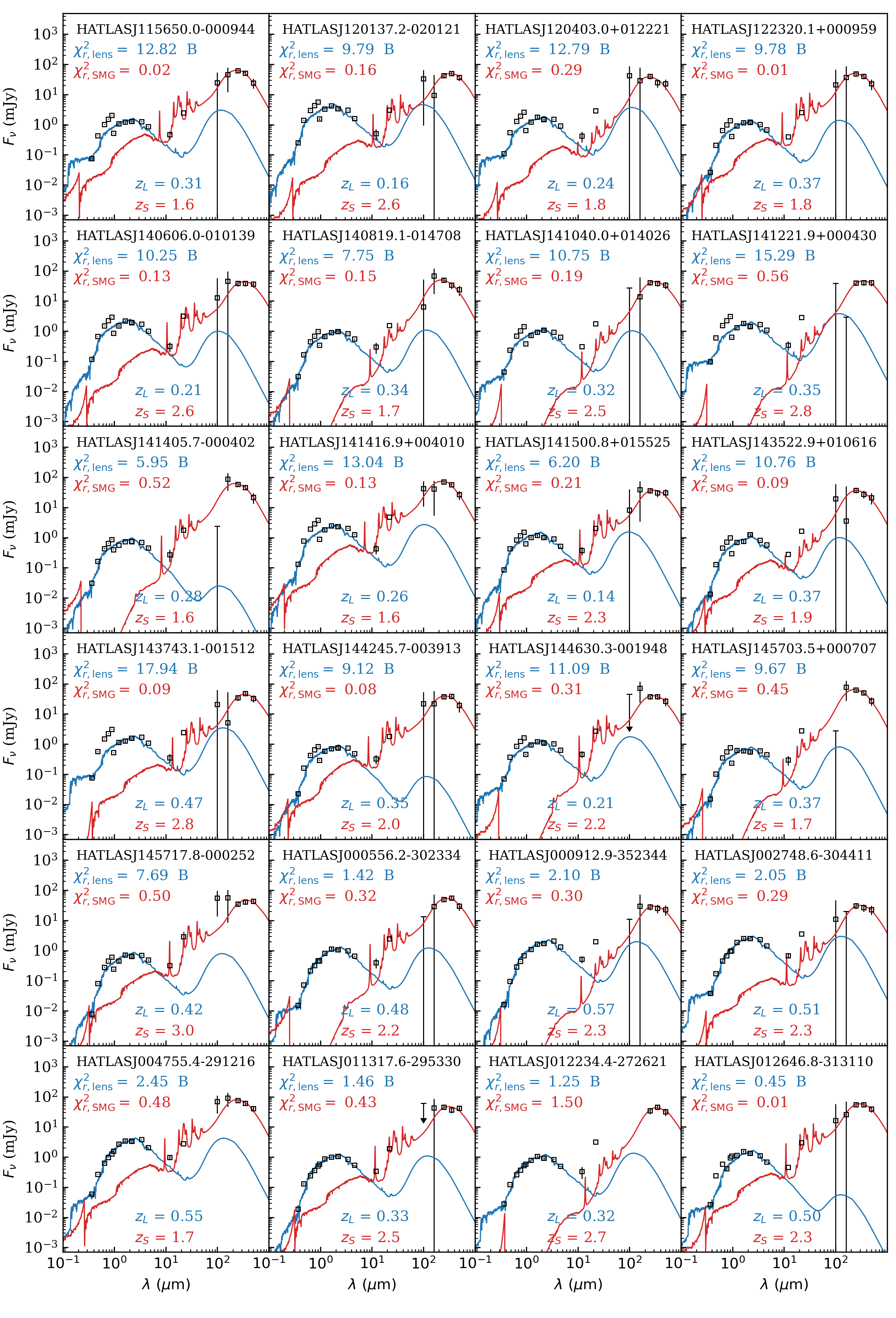}
\caption{continued.}\label{fig:SED_fits_2}
\end{figure*}

\begin{figure*}
\ContinuedFloat
\centering
\includegraphics[width=0.8\textwidth]{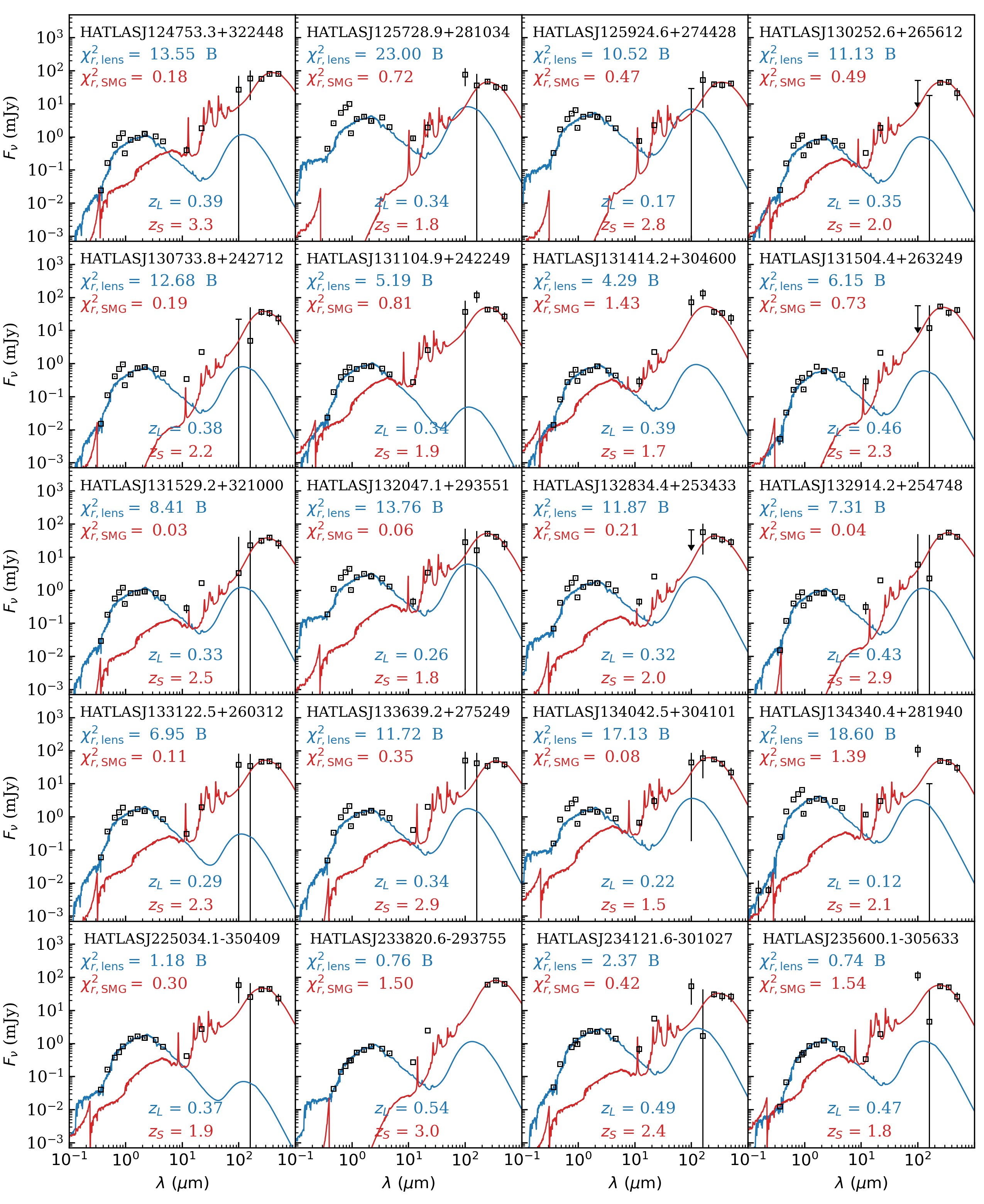}
\caption{continued.}\label{fig:SED_fits_3}
\end{figure*}

\onecolumn
\section{Catalog of candidates}\label{sec:full_catalog}
The 68 candidate lenses and lensed galaxies are listed in Table \ref{tab:list_candidates}, together with their angular separation, an isolation flag in the catalogs used, the reduced chi-squared values for both components of the SEDs, a lensing rank highlighting the most probable lenses, and the photometric redshifts and stellar masses derived from CIGALE. 

{\small
\begin{longtable}{lcccccccc}
\caption{List of candidate lensed galaxies with WISE elliptical counterparts found in H-ATLAS fields.}\label{tab:list_candidates} \\
\hline
\noalign{\smallskip}
HATLAS\_IAU\_ID & $\theta_{H}$ (") & Isolated & $z_s$ & $z_l$ & $\chi^2_{r,\ \text{SMG}}$ & $\chi^2_{r,\ \text{lens}}$ & $\log(M_\star\ /\ M_\odot)$ & Rank \\
\noalign{\smallskip}
\hline
\noalign{\smallskip}
\endfirsthead
\caption{continued.}\\
\hline
\noalign{\smallskip}
HATLAS\_IAU\_ID & $\theta_{H}$ (") & Isolated & $z_s$ & $z_l$ & $\chi^2_{r,\ \text{SMG}}$ & $\chi^2_{r,\ \text{lens}}$ & $\log(M_\star\ /\ M_\odot)$ & Rank \\
\noalign{\smallskip}
\hline
\noalign{\smallskip}
\endhead
\hline
\endfoot
J085752.0$+$012635 & 3.6 & 110 & 2.5 $\pm$ 0.7 & 0.29 $\pm$ 0.04 & 0.29 & 9.05 & 11.35 $\pm$ 0.13 & A \\
J143931.2$+$003339 & 3.2 & 100 & 2.6 $\pm$ 0.7 & 0.33 $\pm$ 0.04 & 0.44 & 7.79 & 11.34 $\pm$ 0.10 & A \\
J144110.6$-$002759 & 4.7 & 100 & 2.1 $\pm$ 0.5 & 0.47 $\pm$ 0.05 & 0.17 & 9.77 & 11.33 $\pm$ 0.09 & A \\
J003647.0$-$282001 & 3.7 & 001 & 1.9 $\pm$ 0.6 & 0.45 $\pm$ 0.05 & 0.29 & 1.11 & 11.68 $\pm$ 0.08 & A \\
J004840.5$-$291409 & 2.4 & 001 & 2.4 $\pm$ 0.6 & 0.31 $\pm$ 0.07 & 0.38 & 1.20 & 11.19 $\pm$ 0.19 & A \\
J005133.5$-$305926 & 3.6 & 000 & 2.6 $\pm$ 0.6 & 0.52 $\pm$ 0.04 & 0.30 & 2.76 & 12.11 $\pm$ 0.08 & A \\
J012654.9$-$300145 & 2.8 & 000 & 2.5 $\pm$ 0.6 & 0.40 $\pm$ 0.05 & 0.37 & 4.66 & 12.17 $\pm$ 0.11 & A \\
J013411.8$-$305646 & 2.4 & 000 & 3.4 $\pm$ 0.5 & 0.51 $\pm$ 0.08 & 0.32 & 1.64 & 11.70 $\pm$ 0.09 & A \\
J014247.5$-$301140 & 3.8 & 000 & 2.1 $\pm$ 0.5 & 0.50 $\pm$ 0.06 & 0.70 & 0.87 & 11.71 $\pm$ 0.08 & A \\
J130800.7$+$292913 & 3.8 & 001 & 2.2 $\pm$ 0.5 & 0.34 $\pm$ 0.04 & 0.06 & 5.15 & 11.28 $\pm$ 0.09 & A \\
J131247.0$+$303802 & 1.6 & 000 & 2.0 $\pm$ 0.6 & 0.33 $\pm$ 0.05 & 0.64 & 5.56 & 11.18 $\pm$ 0.10 & A \\
J131745.1$+$280958 & 2.9 & 001 & 1.9 $\pm$ 0.4 & 0.18 $\pm$ 0.03 & 0.16 & 6.42 & 11.07 $\pm$ 0.14 & A \\
J134135.8$+$284212 & 1.6 & 000 & 1.9 $\pm$ 0.5 & 0.23 $\pm$ 0.05 & 0.18 & 8.88 & 11.15 $\pm$ 0.16 & A \\
J135031.8$+$250444 & 2.0 & 000 & 1.6 $\pm$ 0.4 & 0.37 $\pm$ 0.05 & 1.53 & 4.38 & 11.27 $\pm$ 0.09 & A \\
J232202.1$-$345556 & 1.8 & 000 & 2.1 $\pm$ 0.6 & 0.26 $\pm$ 0.03 & 0.32 & 1.29 & 11.54 $\pm$ 0.10 & A \\
J084406.6$+$015336 & 10.9 & 110 & 2.1 $\pm$ 0.6 & 0.38 $\pm$ 0.06 & 0.01 & 16.14 & 11.92 $\pm$ 0.15 & B \\
J090312.5$-$004509 & 2.4 & 111 & 1.7 $\pm$ 0.5 & 0.17 $\pm$ 0.03 & 0.51 & 14.43 & 11.15 $\pm$ 0.18 & B \\
J090554.8$+$004011 & 13.8 & 110 & 1.5 $\pm$ 0.3 & 0.16 $\pm$ 0.03 & 0.59 & 6.50 & 10.97 $\pm$ 0.15 & B \\
J091249.4$-$000403 & 5.7 & 100 & 2.0 $\pm$ 0.6 & 0.44 $\pm$ 0.02 & 0.83 & 1.86 & 11.95 $\pm$ 0.05 & B \\
J113127.5$+$000317 & 5.5 & 100 & 2.4 $\pm$ 0.7 & 0.32 $\pm$ 0.03 & 1.52 & 7.89 & 11.40 $\pm$ 0.10 & B \\
J113918.8$-$012015 & 3.2 & 101 & 1.6 $\pm$ 0.5 & 0.19 $\pm$ 0.04 & 0.40 & 11.83 & 11.16 $\pm$ 0.18 & B \\
J114411.4$-$015154 & 5.5 & 100 & 1.7 $\pm$ 0.5 & 0.16 $\pm$ 0.05 & 0.24 & 4.56 & 11.10 $\pm$ 0.20 & B \\
J114837.1$-$011246 & 0.4 & 100 & 1.8 $\pm$ 0.5 & 0.25 $\pm$ 0.09 & 0.55 & 14.10 & 11.54 $\pm$ 0.37 & B \\
J115141.3$-$023223 & 13.1 & 111 & 2.0 $\pm$ 0.6 & 0.48 $\pm$ 0.04 & 0.19 & 2.13 & 11.55 $\pm$ 0.08 & B \\
J115650.0$-$000944 & 0.3 & 101 & 2.3 $\pm$ 0.6 & 0.31 $\pm$ 0.09 & 0.02 & 12.82 & 11.37 $\pm$ 0.25 & B \\
J120137.2$-$020121 & 8.2 & 100 & 2.2 $\pm$ 0.6 & 0.16 $\pm$ 0.03 & 0.16 & 9.79 & 11.20 $\pm$ 0.17 & B \\
J120403.0$+$012221 & 7.9 & 100 & 2.2 $\pm$ 0.6 & 0.24 $\pm$ 0.07 & 0.29 & 12.79 & 11.23 $\pm$ 0.28 & B \\
J122320.1$+$000959 & 10.3 & 100 & 2.5 $\pm$ 0.5 & 0.37 $\pm$ 0.04 & 0.01 & 9.78 & 11.47 $\pm$ 0.08 & B \\
J140606.0$-$010139 & 10.2 & 100 & 2.6 $\pm$ 0.7 & 0.21 $\pm$ 0.04 & 0.13 & 10.25 & 11.19 $\pm$ 0.22 & B \\
J140819.1$-$014708 & 6.0 & 100 & 2.3 $\pm$ 0.6 & 0.34 $\pm$ 0.04 & 0.15 & 7.75 & 11.28 $\pm$ 0.11 & B \\
J141040.0$+$014026 & 4.3 & 101 & 2.3 $\pm$ 0.6 & 0.32 $\pm$ 0.05 & 0.19 & 10.75 & 11.27 $\pm$ 0.14 & B \\
J141221.9$+$000430 & 6.8 & 101 & 2.2 $\pm$ 0.4 & 0.35 $\pm$ 0.08 & 0.56 & 15.29 & 11.59 $\pm$ 0.20 & B \\
J141405.7$-$000402 & 7.7 & 110 & 3.7 $\pm$ 0.5 & 0.28 $\pm$ 0.04 & 0.52 & 5.95 & 11.05 $\pm$ 0.13 & B \\
J141416.9$+$004010 & 10.7 & 100 & 1.8 $\pm$ 0.5 & 0.26 $\pm$ 0.07 & 0.13 & 13.04 & 11.45 $\pm$ 0.27 & B \\
J141500.8$+$015525 & 11.3 & 100 & 1.8 $\pm$ 0.4 & 0.14 $\pm$ 0.03 & 0.21 & 6.20 & 10.61 $\pm$ 0.19 & B \\
J143522.9$+$010616 & 6.2 & 100 & 2.4 $\pm$ 0.6 & 0.37 $\pm$ 0.04 & 0.09 & 10.76 & 11.44 $\pm$ 0.08 & B \\
J143743.1$-$001512 & 0.7 & 101 & 1.9 $\pm$ 0.5 & 0.47 $\pm$ 0.07 & 0.09 & 17.94 & 11.77 $\pm$ 0.14 & B \\
J144245.7$-$003913 & 10.2 & 100 & 1.9 $\pm$ 0.5 & 0.35 $\pm$ 0.04 & 0.08 & 9.12 & 11.23 $\pm$ 0.12 & B \\
J144630.3$-$001948 & 5.3 & 101 & 2.3 $\pm$ 0.5 & 0.21 $\pm$ 0.06 & 0.31 & 11.09 & 10.98 $\pm$ 0.29 & B \\
J145703.5$+$000707 & 7.8 & 101 & 2.3 $\pm$ 0.6 & 0.37 $\pm$ 0.04 & 0.45 & 9.67 & 11.23 $\pm$ 0.09 & B \\
J145717.8$-$000252 & 9.3 & 100 & 2.8 $\pm$ 0.5 & 0.42 $\pm$ 0.04 & 0.50 & 7.69 & 11.44 $\pm$ 0.08 & B \\
J000556.2$-$302334 & 11.8 & 001 & 2.0 $\pm$ 0.6 & 0.48 $\pm$ 0.04 & 0.32 & 1.42 & 11.77 $\pm$ 0.07 & B \\
J000912.9$-$352344 & 12.2 & 000 & 2.7 $\pm$ 0.6 & 0.57 $\pm$ 0.04 & 0.30 & 2.10 & 12.19 $\pm$ 0.07 & B \\
J002748.6$-$304411 & 5.7 & 000 & 2.8 $\pm$ 0.7 & 0.51 $\pm$ 0.04 & 0.29 & 2.05 & 12.18 $\pm$ 0.07 & B \\
J004755.4$-$291216 & 5.5 & 000 & 2.2 $\pm$ 0.6 & 0.55 $\pm$ 0.04 & 0.48 & 2.45 & 12.39 $\pm$ 0.07 & B \\
J011317.6$-$295330 & 5.4 & 001 & 2.7 $\pm$ 0.6 & 0.33 $\pm$ 0.07 & 0.43 & 1.46 & 11.49 $\pm$ 0.14 & B \\
J012234.4$-$272621 & 10.0 & 000 & 3.0 $\pm$ 0.7 & 0.32 $\pm$ 0.08 & 1.50 & 1.25 & 11.38 $\pm$ 0.17 & B \\
J012646.8$-$313110 & 7.7 & 001 & 1.7 $\pm$ 0.5 & 0.50 $\pm$ 0.06 & 0.01 & 0.45 & 11.91 $\pm$ 0.10 & B \\
J124753.3$+$322448 & 8.9 & 000 & 2.2 $\pm$ 0.6 & 0.39 $\pm$ 0.05 & 0.18 & 13.55 & 11.47 $\pm$ 0.12 & B \\
J125728.9$+$281034 & 6.3 & 001 & 2.4 $\pm$ 0.6 & 0.34 $\pm$ 0.11 & 0.72 & 23.00 & 11.92 $\pm$ 0.30 & B \\
J125924.6$+$274428 & 9.6 & 001 & 1.9 $\pm$ 0.5 & 0.17 $\pm$ 0.04 & 0.47 & 10.52 & 11.34 $\pm$ 0.20 & B \\
J130252.6$+$265612 & 9.3 & 001 & 2.8 $\pm$ 0.7 & 0.35 $\pm$ 0.04 & 0.49 & 11.13 & 11.26 $\pm$ 0.11 & B \\
J130733.8$+$242712 & 10.4 & 000 & 2.2 $\pm$ 0.6 & 0.38 $\pm$ 0.05 & 0.19 & 12.68 & 11.28 $\pm$ 0.12 & B \\
J131104.9$+$242249 & 10.7 & 000 & 2.5 $\pm$ 0.6 & 0.34 $\pm$ 0.04 & 0.81 & 5.19 & 11.29 $\pm$ 0.09 & B \\
J131414.2$+$304600 & 5.5 & 000 & 1.4 $\pm$ 0.4 & 0.39 $\pm$ 0.04 & 1.43 & 4.29 & 11.37 $\pm$ 0.07 & B \\
J131504.4$+$263249 & 7.2 & 000 & 2.6 $\pm$ 0.6 & 0.46 $\pm$ 0.05 & 0.73 & 6.15 & 11.48 $\pm$ 0.08 & B \\
J131529.2$+$321000 & 10.9 & 001 & 1.8 $\pm$ 0.6 & 0.33 $\pm$ 0.03 & 0.03 & 8.41 & 11.30 $\pm$ 0.10 & B \\
J132047.1$+$293551 & 8.3 & 000 & 2.4 $\pm$ 0.6 & 0.26 $\pm$ 0.06 & 0.06 & 13.76 & 11.50 $\pm$ 0.21 & B \\
J132834.4$+$253433 & 5.5 & 000 & 2.1 $\pm$ 0.6 & 0.32 $\pm$ 0.04 & 0.21 & 11.87 & 11.49 $\pm$ 0.13 & B \\
J132914.2$+$254748 & 12.3 & 001 & 2.4 $\pm$ 0.6 & 0.43 $\pm$ 0.03 & 0.04 & 7.31 & 11.50 $\pm$ 0.07 & B \\
J133122.5$+$260312 & 14.1 & 000 & 2.6 $\pm$ 0.5 & 0.29 $\pm$ 0.04 & 0.11 & 6.95 & 11.41 $\pm$ 0.11 & B \\
J133639.2$+$275249 & 8.6 & 001 & 2.9 $\pm$ 0.5 & 0.34 $\pm$ 0.04 & 0.35 & 11.72 & 11.48 $\pm$ 0.11 & B \\
J134042.5$+$304101 & 5.5 & 000 & 2.6 $\pm$ 0.8 & 0.22 $\pm$ 0.05 & 0.08 & 17.13 & 11.12 $\pm$ 0.26 & B \\
J134340.4$+$281940 & 8.9 & 000 & 2.1 $\pm$ 0.5 & 0.12 $\pm$ 0.03 & 1.39 & 18.60 & 11.07 $\pm$ 0.21 & B \\
J225034.1$-$350409 & 6.8 & 000 & 2.6 $\pm$ 0.6 & 0.37 $\pm$ 0.09 & 0.30 & 1.18 & 11.67 $\pm$ 0.16 & B \\
J233820.6$-$293755 & 5.1 & 001 & 1.4 $\pm$ 0.3 & 0.54 $\pm$ 0.06 & 1.50 & 0.76 & 11.72 $\pm$ 0.09 & B \\
J234121.6$-$301027 & 5.2 & 000 & 2.0 $\pm$ 0.6 & 0.49 $\pm$ 0.05 & 0.42 & 2.37 & 12.09 $\pm$ 0.11 & B \\
J235600.1$-$305633 & 5.3 & 000 & 1.5 $\pm$ 0.4 & 0.47 $\pm$ 0.06 & 1.54 & 0.74 & 11.81 $\pm$ 0.09 & B \\
\noalign{\smallskip}
\hline
\end{longtable}
\tablefoot{For each source, the following columns represent the angular separation of the candidate lenses with respect to H-ATLAS sources ($\theta_H$); an isolation flag for the foreground galaxies (where 1 means ``isolated'', and 0 that there are one or more counterparts; first digit for optical wavelengths, second for NIR, and third for WISE); the photometric redshift for the SMGs and lenses as derived by CIGALE ($z_s$ and $z_l$, respectively); the goodness of fit of the models ($\chi^2_{r,\ \text{SMG}}$, $\chi^2_{r,\ \text{lens}}$); the logarithm of the stellar mass of the lenses ($\log(M_\star\ /\ M_\odot)$); and a lensing rank indicating the reliability of the lenses (A=best candidates, B=other candidates).}
}
\twocolumn

\end{appendix}
\end{document}